\documentclass[11pt]{article}
\usepackage{jheppubmodnew}
\usepackage{float}
\pdfoutput=1
\usepackage{amssymb}
\usepackage{amsmath}
\usepackage{amsthm}
\usepackage{mathtools}
\usepackage{graphicx}
\usepackage{xcolor}
\usepackage{simpler-wick}
\usepackage[labelsep=quad]{subcaption}
\usepackage{cite}
\usepackage{epsfig}
\usepackage{braket}
\usepackage[colorlinks=true]{hyperref}
\usepackage{simpler-wick}

\makeatletter
\newcommand{\llangle}{\mathopen{\langle\!\langle}}
\newcommand{\rrangle}{\mathclose{\rangle\!\rangle}}
\makeatother

\newcommand{\dbar}{{\overline{\Delta}}}
\newcommand{\dsvol}{\text{vol(SO(1,$d+1$))}}

\newcommand{\rotgrpvol}{\text{vol(SO($d-1$))}}

\newcommand{\bchi}{\overline{\chi}}
\newcommand{\bJ}{\overline{J}}

\newcommand{\G}{\Gamma}
\newcommand{\lm}{\lambda}

\newcommand{\g}{\delta}

\usepackage{mathrsfs}
\newcommand{\D}{\Delta}

\newcommand{\ra}{\rangle}

\newcommand{\psing}{\Psi} 
\newcommand{\psiobs}{\Psi_{\text{obs}}}
\newcommand{\alset}{{\cal A}_{\text{light}}}
\newcommand{\psiback}{\Psi_{\text{back}}}
\newcommand{\psibacksm}{\psi_{\text{back}}}
\newcommand{\mcalc}{C}
\newcommand{\qftn}{} \newcommand{\normconst}{{\cal K}}
\newcommand{\fourpw}{\Theta}

\def\pb[#1,#2]{\{#1, #2\}}
\def\deb[#1,#2]{[#1,#2]_{\text{D.B.}}}

\def\a{\alpha}

\def\Or[#1]{{\text{O}}\left({#1}\right)}
\def\dotl[#1,#2]{\left\langle #1,\, #2 \right\rangle}
\def\dotlb[#1,#2]{\left\langle #1,\, #2 \right\rangle}
\def\dotlm[#1,#2]{\left[ #1,\, #2 \right]}
\def\dotp[#1,#2]{(\vect{#1} \cdot\vect{#2})}
\def\aff[#1,#2]{\hat{#1}(#2)}

\def\n4sym{{\cal N}=4 SYM}
\def\>{\rangle}
\def\<{\langle}
\def\weight[#1,#2,#3]{\{(#1),#2,#3\}}
\def\ads[#1]{$\text{AdS}_{#1}$}

\newcommand{\be}{\begin{equation}}
\newcommand{\ee}{\end{equation}}
\newcommand{\ba}{\begin{align}}
\newcommand{\ea}{\end{align}}
\newcommand{\bs}{\begin{split}}
\def\sess\end{split}
\newcommand{\vect}[1]{{\boldsymbol{#1}}}

\def \bea {\begin{eqnarray}}
\def \eea {\end{eqnarray}}
\def \bea* {\begin{eqnarray*}}
\def \eea* {\end{eqnarray*}}
\def \bes {\begin{equation*}}
\def \ees {\end{equation*}}

\def \lm  {\lambda}
\def \b  {\beta}

\def\confgrp{{\mathrm{SO}}(1,d+1)}
\def\sodminone{{\mathrm{SO}}(d-1)}

\def\scoeff[#1]{{\mathcal G}_{#1}}
\def\coeff[#1,#2]{{\mathcal G}_{#1,#2}}
\def\dcoeff[#1]{\delta G_{#1}} \def\tcoeff[#1,#2]{{\widetilde{\mathcal G}}_{#1,#2}}
\def\lcoeff[#1,#2]{{\mathcal G}^{\lambda}_{#1,#2}}

\def\cprod{{\cal C}}

\def\alcut[#1]{{\cal A}_{#1, \epsilon}}
\def\alseg[#1,#2]{{\cal B}_{#1, #2}}

\def\supcharge[#1]{\{#1\}}

\def\projsupeig[#1]{{\cal P}_{{\ell, m}}[{#1}]}
\def\transop[#1, #2]{T_{\{#1\}, \{#2\}}}
\def\supket[#1]{|\{#1\} \rangle}
\def\supbra[#1]{\langle \{#1\} | }

\def\hilbzerosect[#1]{{\cal H}^0_{\{#1\}}}
\def\hilbsect[#1]{{\cal H}_{\{#1\}}}
\def\hilbmass[#1]{{\cal H}^m}

\newcommand{\x}{\chi}

\newcommand{\enc}[1]{\left( #1\right)}

\newcommand{\encbr}[1]{\left\{#1\right\}}

\DeclarePairedDelimiter\abs{\lvert}{\rvert}

\def\vol{\mathrm{vol}}

\renewcommand{\b}{\beta}
\renewcommand{\a}{\alpha}

\renewcommand{\G}{\Gamma}

\begin{document}

\title{Cosmological correlators in gravitationally-constrained de Sitter states}
\author[a,b]{Tuneer Chakraborty,}
\author[a]{Ashik H}
\author[a]{and Suvrat Raju}
\affiliation[a]{International Centre for Theoretical Sciences, Tata Institute of Fundamental Research, Shivakote,  Bengaluru 560089, India}
\affiliation[b]{Department of Theoretical Physics, Tata Institute of Fundamental Research, Homi Bhabha Road, Mumbai 400005, India}
\date{}
\abstract{We study cosmological correlators in de Sitter quantum gravity in the limit where $G_N \to 0$. This limit is distinct from a nongravitational QFT because the gravitational constraints still force states and observables to be de Sitter invariant.  We first examine a class of perturbative correlators that, in gauge-fixed form, are represented by the expectation value of a product of elementary fields on the late-time boundary.   We formulate Feynman rules for our computations and enumerate some necessary, but not sufficient, conditions that must be imposed on states and operators to avoid group-volume divergences. These correlators are conformally invariant in all allowed perturbative states but never coincide with QFT vacuum-expectation values.  For instance, our sample computations yield interesting non-Gaussianities even when the underlying vacuum wavefunction is Gaussian.  However, we show that, in the presence of a heavy background state, it is possible to construct a separate class of state-dependent relational observables whose values approximate QFT correlators in the vacuum. This illustrates a key contrast in quantum gravity ---  between observables that are microscopically simple  and observables whose expectation values in an appropriate background state lead to simple QFT-like correlators.}

\setcounter{tocdepth}{1}

\maketitle
\section{Introduction}
The study of quantum gravity in asymptotically de Sitter (dS) space is of theoretical interest. The AdS/CFT correspondence \cite{Maldacena:1997re,Witten:1998qj,Gubser:1998bc} has revealed many interesting aspects of gravity in anti-de Sitter space; we would like to understand which of these results can be adapted to the de Sitter context and also study the qualitatively new effects that arise only in de Sitter space. From a different perspective, the early Universe is believed to have undergone a period of accelerated expansion when it was well-approximated by de Sitter space. Therefore, a study of de Sitter quantum gravity might be relevant for inflationary physics.

Nevertheless, while quantum field theory in de Sitter space has been studied extensively, and already provides a rich theoretical playground, the study of gravity \cite{Anninos:2012qw} has received less attention. New physical effects in gravity arise from the gravitational constraints. Since our intuition is commonly based on classical solutions of the equations of motion, the significance of the constraints in the quantum theory is often overlooked.  But the most interesting aspects of gravity in asymptotically anti-de Sitter space and asymptotically flat space --- the fact that it is holographic in the sense that all observables in the bulk can also be represented near the asymptotic boundary --- can be derived from a careful study of the quantum-mechanical gravitational constraints \cite{Marolf:2013iba,Laddha:2020kvp,Raju:2020smc}.

The papers \cite{Chakraborty:2023yed,Chakraborty:2023los} studied the effects of the quantum gravitational constraints in asymptotically de Sitter space by classifying all solutions of the Wheeler-DeWitt equation. These papers also proposed a norm on this space to obtain a Hilbert space. Finally, they introduced a class of gauge-fixed observables, termed ``cosmological correlators'', that are  labelled by points on the asymptotically late-time slice, although their gauge-invariant description involves an integral on the entire late-time slice. These observables are the natural microscopically simple observables in a theory of quantum gravity in asymptotically de Sitter space and amenable to straightforward perturbative computations. In this paper, we will continue to use the phrase ``cosmological correlators'' to describe them although we comment more on the terminology below. 

These observables obey a surprising property:  cosmological correlators labelled by points in any open set of the late-time slice form a complete set of observables. This result can be viewed as an extension of the principle of holography of information \cite{Laddha:2020kvp,Chowdhury:2020hse,Raju:2020smc,Chowdhury:2021nxw,Raju:2021lwh,Chakravarty:2023cll,deMelloKoch:2022sul} to de Sitter space. We use the phrase ``holography of information'' because this result, which is related to similar results in anti-de Sitter space and flat space, tells us that quantum information about the entire spacetime is available in a restricted region but does not provide a full-fledged holographic dual that captures the dynamics of the theory. The natural hope for a full holographic dual would be to show, by some extension of these arguments, that dynamics on all of de Sitter space can be captured by degrees of freedom on a codimension-1 region of the late-time slice.

In this paper, we advance this program by explicitly studying the norm and cosmological correlators in the limit where $G_N \to 0$ but the fields continue interacting via perturbative nongravitational interactions. One might naively expect that, in the limit $G_N \to 0$, all gravitational effects vanish. However, this is incorrect; in \cite{Chakraborty:2023yed,Chakraborty:2023los}, it was shown that the $G_N \to 0$ limit of the gravitational Hilbert space is not the QFT Hilbert space. Physically, this is because as one takes $G_N \to 0$, it is necessary to still impose the Gauss law \cite{moncrief1978invariant,Higuchi:1991tk,Higuchi:1991tm,Marolf:2008hg,Marolf:2008it,Anninos:2017eib}. The Gauss law restricts the Hilbert space by imposing de Sitter invariance on all states and observables.  

We find the following results
\begin{enumerate}
\item
As explained in \cite{Chakraborty:2023yed,Chakraborty:2023los} the Euclidean vacuum (also known as the Hartle-Hawking state \cite{Hartle:1983ai}) which we denote by $|0 \rangle$, is not perturbatively normalizable.  A  necessary condition for excited states to be normalizable is that they should be orthogonal to $|0 \rangle$. We formulate Feynman rules for computing the norm of such states in perturbation theory and provide some sample computations.
\item
We formulate Feynman rules for computing cosmological correlators and provide sample computations of these observables. We formulate necessary conditions for such computations to give finite values that are free of group-volume divergences.
\item
These cosmological correlators always differ from QFT-correlators in the vacuum. For instance, they can display interesting non-Gaussian features even when the vacuum corresponds to a  Gaussian wavefunction. We present several explicit examples in this paper. There is nothing mysterious about this because (a) every normalizable state has excitations on top of the vacuum (b) the gauge-fixed operators that we study are labelled by points but are secretly smeared over the entire late-time slice. Therefore, their correlators pick up the presence of the excitation. One noteworthy difference  from QFT correlators in excited states is that these correlators are always conformally invariant.
\item
Nevertheless, in the presence of an appropriate background state, it is possible to define ``relational observables'' whose expectation values resemble QFT correlators in the vacuum. This construction generalizes the AdS construction of relational observables presented in \cite{Papadodimas:2015xma,Papadodimas:2015jra,Bahiru:2022oas,Bahiru:2023zlc,Jensen:2024dnl}.  Here, the microscopic description of both the state and the observables is complicated but the final expectation values correspond to simple QFT observables.

The background state can be thought of as an ``observer'' as discussed in the recent literature \cite{Chandrasekaran:2022cip}. However, the ``observer'' is not introduced by hand as an auxiliary tool to solve the constraints.   Instead, we study an excitation about the vacuum and set up relational observables while manifestly respecting all the original constraints.
\end{enumerate}

These results buttress the broad narrative that has developed over a number of studies. The gravitational constraints allow one to deduce a version of the holographic principle; this can be explicitly verified for simple perturbative states and observables; nevertheless, it is possible to construct states and observables where our usual notions of locality in quantum field theory are restored to good approximation. 

We believe that our perturbative computations are of theoretical interest in understanding de Sitter-quantum gravity. However, we do not suggest that they have any immediate observational consequence. It is more natural that the correlations we observe in the cosmic microwave background or the large-scale structure of the Universe correspond to correlators of the  ``relational observables'' described above, which are close to standard QFT observables.  To avoid confusion, we do not use the phrase ``cosmological correlators'' for correlators of these relational observables in this paper. 

These relational observables themselves deserve further study --- both from a technical and a fundamental perspective. An important technical problem is to develop a systematic perturbative treatment of these relational observables going beyond the leading-order treatment that we provide here.  A more conceptual issue is to understand their ``state dependence'', which is similar to the state dependence of observables \cite{Papadodimas:2013wnh,Papadodimas:2013jku} in the black-hole interior and raises deeper questions about observations in a cosmological setting, where the observer is part of the Universe they are studying.

A brief overview of this paper is as follows. In Section \ref{secsetup}, we review our setup, including key results from \cite{Chakraborty:2023yed,Chakraborty:2023los}. In Section \ref{secfeynman}, we set up Feynman rules to compute perturbative observables in the nongravitational limit. Section \ref{secsample} presents several sample calculations. Section \ref{secrelational} describes the construction of observables dressed to an observer whose values correspond to QFT vacuum correlators. Appendix \ref{seccosmgaugeinv} shows how to find the gauge-invariant operator corresponding to a gauge-fixed cosmological correlator. Appendix \ref{apprelationaldetails} provides further details of the relational construction and also includes a quick review of the corresponding construction in AdS/CFT that might be useful for some readers. Appendix \ref{appprincipalwavefunc} explains the wavefunction formalism for principal-series fields, which is slightly subtle because these wavefunctions must be viewed as coherent-state wavefunctions.

\section{Review \label{secsetup}}
The papers \cite{Chakraborty:2023yed,Chakraborty:2023los} constrained the allowed states in a theory of quantum gravity in asymptotically de Sitter space by finding all possible solutions of the Wheeler-DeWitt equation on the asymptotically late-time slice. The solution space was lifted to a Hilbert space by means of an appropriate norm and a version of the principle of holography of information was proved in terms of gauge-fixed observables termed cosmological correlators. In this section, we provide a self-contained review of the relevant results from \cite{Chakraborty:2023yed,Chakraborty:2023los}. 

In this paper, our first objective is to explore the structure of the norm and gauge-fixed cosmological correlators defined in \cite{Chakraborty:2023yed,Chakraborty:2023los}  within perturbation theory. For simplicity, we will focus on the $G_N \to 0$ limit but allow nongravitational interactions to persist. This limit was studied in sections 5.2 and 5.3 of \cite{Chakraborty:2023yed} and section 5.1 of \cite{Chakraborty:2023los} and our review summarizes those sections. 

We focus on $d+1$ dimensional asymptotically de Sitter spacetimes where $d \geq 2$. However, interesting and related results have been obtained by studying the Wheeler-DeWitt equation in $1+1$ dimensions \cite{Henneaux:1985nw,Nanda:2023wne,Godet:2024ich,Parrikar:2025xmz}.

\paragraph{\bf Matter fields and boundary values.}
We use massive scalar fields to model the matter-sector of the theory. Consider a late-time slice of de Sitter space where the spatial metric takes the form
\be
g_{i j} \underset{\Omega \to \infty}{\longrightarrow} {\Omega^2 \over (1 + |x|^2)^2} (\delta_{i j} + \kappa h_{i j}),
\ee
where $\kappa = \sqrt{8 \pi G_N}$, which is the metric of a sphere with small fluctuations scaled up by a large Weyl factor. Since we focus on the $G_N \to 0$ limit, we will ignore the effect of the metric fluctuations $h_{i j}$ in this paper. A scalar field of mass $m$ is expected to die off in the large $\Omega$ limit as
\be
\label{latetimefalloff}
\chi \underset{\Omega \to \infty}{\longrightarrow} \left({\Omega \over 1 + |x|^2}\right)^{-\Delta} \chi(x) + \left({\Omega \over 1 + |x|^2}\right)^{-\dbar} \bchi(x)  + \ldots,
\ee
where $\Delta(d-\Delta) = m^2$ in the free-field limit.  This is simply the usual ``extrapolate limit'' in de Sitter space.\footnote{In the interacting theory, the map \eqref{latetimefalloff} between bulk and boundary fields might need to be improved as discussed in \cite{SalehiVaziri:2024joi}.}
We denote $\dbar \equiv d - \Delta$. For light fields $\Delta$ is real and can be chosen so that $\Delta < {d \over 2}$. In this case, the fields are said to belong to the ``complementary series'' ;the first term displayed in \eqref{latetimefalloff} is the dominant term at late times and $\chi(x)$ can be taken to be real.   Fields with larger values of $m$ correspond to complex values of $\Delta = {d \over 2} + i \mu$ and $\dbar = {d \over 2} - i \mu$. In this case, the fields are said to belong to the ``principal series.''  In that case the second term in  \eqref{latetimefalloff} is also important.

We will present expressions below where we distinguish between $\chi$ and $\bchi$ so that they are valid when $\chi$ is in the principal series. For complementary-series fields, the reader should simply identify $\chi$ and $\bchi$.

\paragraph{\bf States.}
Let $|0 \rangle$ be the Euclidean vacuum. We use the term ``Euclidean vacuum'' interchangeable with ``Hartle-Hawking state''.

In the nongravitational limit, it was shown in \cite{Chakraborty:2023yed} that the states satisfying the Gauss law take the form
\be
\label{psingdef}
|\psing \rangle =  \sum_{m} \int d \vec{x} \,  \dcoeff[m]( \vec{x})  \bchi(x_1) \ldots \bchi(x_m)  |0 \rangle,
\ee
where $\vec{x} \equiv (x_1, \ldots, x_m)$ and  $\dcoeff[m](\vec{x})$ is a function that obeys the same Ward identities as the connected correlation function of $m$ operators of dimension $\Delta$. 
\be
\dcoeff[m](\vec{x}) \sim \langle O(x_1) \ldots O(x_m) \rangle_{\text{connected}}.
\ee
We omit the subscript $\text{ng}$ on the states that was used in \cite{Chakraborty:2023yed,Chakraborty:2023los} since we are always in the nongravitational limit.

We would like to make a few comments. 
First, as we explain in Appendix \ref{appprincipalwavefunc} it is convenient to think of states and excitations produced by excitations of $\bchi$ rather than $\x$. Of course, when $\x$ is in the complementary series, this distinction is irrelevant. 
Second, we use $\sim$ rather than $=$ to emphasize that the function on the left is only constrained by the Ward identities but need not necessarily obey the constraints of locality or unitarity that such correlators would obey in a physical CFT.  The subscript ``connected'' also implies that $\dcoeff[m](\vec{x})$ is not invariant under separate conformal transformations of subsets of its arguments, and this will be important later.  
Third, on the late-time slice that we are studying, the action of the de Sitter group reduces to the conformal group. It is easy to see that the states \eqref{psingdef} are invariant under conformal transformations: the transformation of $\dcoeff[m]$ under a conformal transformation is compensated by the transformation of $\bchi(x_i)$. It is also easy to see that these states correspond to the group-averaged Hilbert space described by Higuchi \cite{Higuchi:1991tk,Higuchi:1991tm}. We refer the reader to \cite{Chakraborty:2023yed}  for more details.

The form \eqref{psingdef} was derived in  \cite{Chakraborty:2023yed,Chakraborty:2023los} from a detailed analysis of a nongravitational limit of solutions of the Wheeler-DeWitt equation in asymptotically de Sitter space. Nevertheless, the  invariance of states under the conformal group can be understood via a simple physical argument.  When one studies a gauge theory in a noncompact space, the Hilbert space contains states that carry nonzero charge and therefore  transform in nontrivial representations of the global symmetry group. In the same way, when one studies gravity in asymptotically flat space or anti-de Sitter space, it is natural to include states that have nonzero charges under the asymptotic symmetry group. However, when one places the gauge theory on a compact manifold, the Gauss law forces us to restrict attention to states that transform in the trivial representation of the symmetry group. For instance, this has interesting physical consequences when one studies SU(N) theories on a sphere \cite{Aharony:2003sx}. This constraint must be preserved even as one takes the gauge coupling to zero. In global de Sitter space, the Cauchy slices are compact. Therefore, the gravitational Gauss law forces us to consider de Sitter invariant states. We comment more on the contrast between the classical and quantum case below, after discussing the norm.

\paragraph{\bf Norm.}
The  states \eqref{psingdef} are not normalizable in the standard QFT norm. This is clear because the smearing functions are not compactly supported. So a naive QFT computation of the norm of \eqref{psingdef} receives an infinite contribution coming from the volume of the conformal group. In \cite{Chakraborty:2023los} it was shown that, starting with a natural norm in the gravitational case (obtained by squaring and integrating the wavefunction and dividing by the volume of the  diff and Weyl redundancy) the nongravitational norm reduces to
\be
\left( \psing, \psing  \right) = \normconst \langle \psing | \psing \rangle_{\qftn},
\ee
where, on the right hand side, we simply have the usual QFT norm of the state  given formally by
\be
\langle \Psi | \Psi \rangle_{\qftn}
=  \sum_{m,m'} \int d \vec{x} d \vec{x}' \, \dcoeff[m](\vec{x})  \dcoeff[m']^*(\vec{x}') \langle 0 |   \chi(x'_1) \ldots \chi(x'_m)\bchi(x_1) \ldots\bchi(x_{m'})| 0 \rangle_{\qftn}~.
\ee
The usual angular bracket norm is used exclusively for the QFT norm and we omit the subscript ${\text{QFT}}$ that was used in \cite{Chakraborty:2023yed,Chakraborty:2023los}.

This expression has a divergence that is proportional to the volume of the conformal group that arises from the integrals over the coordinates that appear in the state above. This is cancelled by the constant
\be
{\cal K} \equiv {\text{vol}(\sodminone) \over \text{vol}(\confgrp)}.
\ee
This norm coincides with Higuchi's expression \cite{Higuchi:1991tk,Higuchi:1991tm} up to an unimportant normalization.

One can make sense of the infinite volume of the conformal group simply by fixing three points in the integral above to 0,1, $\infty$. 
\be
\label{normfixed}
\begin{split}
\left( \psing, \psing  \right) = &\sum_{m,m'} \int d \vec{x} d \vec{x}' \delta(z_1) \delta(z_2 - 1) \delta(\tilde{z}_3)  \, \dcoeff[m](\vec{x})  \dcoeff[m']^*(\vec{x}') \\ &\times \langle 0 |   \chi(x'_1) \ldots \chi(x'_{m'})\bchi(x_1) \ldots\bchi(x_{m}) | 0 \rangle,
\end{split}
\ee
where $\tilde{z}_3^{\mu} = {z_3^{\mu} \over |z_3|^2}$,  and $z_1, z_2, z_3$ are {\em any} three coordinates from the set $(x_1, \ldots x_m, x_1', \ldots x'_{m'})$. 
Since the norm requires us to fix at least three points, we note that the vacuum state itself is not a normalizable element of the Hilbert space. This issue has been noted earlier \cite{Higuchi:1991tm,Chandrasekaran:2022cip} and we refer the reader to \cite{Cotler:2025gui} for recent discussion. Higher-genus effects are studied in \cite{Godet:2025bju,Collier:2025lux}.

\paragraph{\bf An aside on dS invariance.}
The construction of dS-invariant states and the norm described above might appear counterintuitive to some readers, particularly those who study classical de Sitter solutions. It is easy to find classical solutions, which satisfy the constraints, but are not de Sitter invariant even asymptotically; so why do the constraints force de Sitter invariance on quantum states? 

A simple analogy might help to explain the difference between the classical and quantum theory. Consider an ordinary quantum field theory in d+1 dimensions, with a local field $\phi(x)$, and let us impose the constraint $P = 0$, where $P$ is the momentum operator. (We have invented this toy model to illustrate some aspects of the de Sitter constraints and we do not claim that it has any deep physical significance.)

In the classical theory it is easy to find solutions that satisfy this constraint. One simply considers a solution to the equation of motion whose center-of-mass momentum is zero. For example, we take one lump of field moving to the left and another moving to the right. 

On the other hand, the quantum constraint $P |\Psi \rangle = 0$ forces states to be translationally invariant. The only normalizable state that satisfies this is the vacuum, $|0 \rangle$. But it is possible to consider states
\be
|\Psi \rangle = \int d x_1 \ldots dx_n \, f(x_1, \ldots x_n) \phi(x_1) \ldots \phi(x_n) | 0 \rangle,
\ee
for any smearing function that satisfies $f(x_1, \ldots x_n) = f(x_1 + a, \ldots x_n + a)$ for all vectors a. These states satisfy the constraint. They are not normalizable in the original QFT norm but one can define a new norm
\be
\left( \Psi, \Psi \right) = {\langle \Psi | \Psi \rangle \over \text{vol}(\mathbb{R}^d)},
\ee
that makes these states normalizable subject to some additional constraints on $f$.  The original vacuum is no longer normalizable although the Hilbert space is again infinite dimensional. We expect that carefully taking the classical limit in this new Hilbert space should lead us back to the set of classical solutions with zero center-of-mass momentum.

\paragraph{\bf Cosmological correlators.} 
The authors of \cite{Chakraborty:2023los}  defined a set of gauge-fixed observables in the constrained de Sitter Hilbert space that were called ``cosmological correlators.'' The definition in the full gravitational theory is given in \cite{Chakraborty:2023los}. In the nongravitational limit, the definition simplifies significantly. The cosmological correlator of n insertions is simply defined by 
\be
\label{ccexpr}
\llangle \psing | \cprod(y_1, \ldots y_n)  | \psing \rrangle_{\text{CC}} =  \langle \psing | \chi(y_1) \ldots \chi(y_n) | \psing \rangle_{\qftn}.
\ee
Note that in the expression above, the points $y_i$ are fixed {\em by hand} and not integrated over.  Of course, it is possible to study a linear combination of such cosmological correlators by smearing them with a sequence of functions $g_{q}(\vec{y})$.  
\be
\label{ccgeneral}
\llangle \psing | \cprod[g_1, \ldots g_n] | \psing \rrangle = \sum_{q=1}^{n} \int d \vec{y} g_{q}(\vec{y}) \langle \psing | \chi(y_1) \ldots \chi(y_q) | \psing \rangle_{\qftn}.
\ee
This is just the usual smearing that is often useful for correlation functions and $g_q(\vec{y})$ are not constrained by conformal invariance.

The expression \eqref{ccgeneral} reduces to a linear combination of QFT vacuum expectation values.
\be
\label{generalcosmo}
\begin{split}
\llangle \psing | \cprod[g_1, \ldots g_n] | \psing \rrangle_{\text{CC}} &= \sum_{m,m',q} \int  d \vec{x} d \vec{x}' d \vec{y} \, \dcoeff[m']^*(\vec{x}')  \dcoeff[m](\vec{x}) g_{q}(\vec{y}) \\ &\times \langle 0 | \chi(x'_1) \ldots \chi(x'_{m'})  \chi(y_1) \ldots \chi(y_q) \bchi(x_1) \ldots\bchi(x_{m})| 0 \rangle_{\qftn}.
\end{split}
\ee

Two comments are in order. First, the reader might worry that the expectation values on the right hand side of \eqref{ccgeneral} suffers from group-volume divergences. Intuitively, 3 fixed points are sufficient to absorb the divergence of the group volume. A little more thought shows that what we really need is that only {\em connected} diagrams with at least three fixed points should contribute to \eqref{generalcosmo}. Following this intuition, it will be shown below that for $n \geq 3$, provided we impose some constraints on the smearing functions $g_q$, the cosmological correlator is finite. 

Second, it was shown in \cite{Chakraborty:2023los} that the cosmological correlators can be written as expectation values of a gauge invariant operator
\be
\llangle \psing | \cprod(y_1, \ldots y_q) | \psing \rrangle_{\text{CC}} = \left(\psing,  O \psing \right),
\ee
where 
\be
\label{angcosm}
O =  {1 \over \text{vol(SO(d-1))}} \int d U U^{\dagger} \chi(y_1) \ldots \chi(y_q)  U,
\ee
and $d U$ is the Haar measure on the conformal group. It is shown in the Appendix that this integral can be performed explicitly in some cases and leads to
\be
O = \int d \vec{z} \, G( \vec{z})  \chi(z_1) \ldots \chi(z_q).
\ee
Here,  $G$ is a function that obeys the Ward identities of CFT correlators in terms of the variables $z_i$.  The function depends on the fixed points  $y_i$ but this dependence is kept implicit in the expression above. Appendix \ref{seccosmgaugeinv} provides explicit examples of this relation. Although our computations will not use the representation \eqref{angcosm}, it is important from a conceptual perspective since it tells us that cosmological correlators are secretly nonlocal observables even though they are labeled by points. This should not be surprising given that a theory of gravity can have no local gauge-invariant observables.

\paragraph{\bf Holography of information.}
In \cite{Chakraborty:2023los}, it was shown that specifying cosmological correlators in any open set of the late-time slice specifies them everywhere.  This is indicative of the holographic properties of gravity: in a nongravitational field theory, correlators on different parts of the late-time slice carry separate pieces of quantum information; but, in gravity, this quantum information is squeezed down to correlators in an infinitesimal region. Incidentally, this is consistent with the conclusion reached in \cite{Bousso:2022hlz} from very different considerations.

As an aside, the phrase ``dS/CFT'' is sometimes used to simply refer to the fact that the coefficients of the Hartle-Hawking wavefunction are related to CFT correlators.  However, this is not necessarily related to holography or even to gravity. The ``CFT'' that appears in this context is simply a machine for determining properties of the Euclidean vacuum. Holography is always a statement about the unusual localization of quantum information in quantum gravity in all states, and not just about the properties of one state.

Our results for cosmological correlators in Section \ref{secsample} serve to verify the holographic properties of gravity in de Sitter space in the perturbative limit.  However, our construction of Section \ref{secrelational} also shows that, in an appropriate limit, holography can be obscured in gravity. This is important since it is by obscuring holography that we regain the QFT limit and mundane notions of locality.

\paragraph{Generating function.}
Before concluding this section, it is useful to introduce some notation that will be helpful below.

We see above that both the norm \eqref{normfixed} and cosmological correlators \eqref{generalcosmo} can be represented as integrated quantum-field-theory correlators in the Euclidean vacuum. So it is convenient to introduce a generating function for these QFT correlators
\be
\begin{split}
&Z[J, \bJ] =   \langle 0 | e^{\int J(x') \chi(x') d x} e^{ \int \bJ(x) \bchi(x) d x} |0 \rangle_{\qftn} \\ &=\sum_{n} {1 \over n! m!}   \int d \vec{x} d \vec{x}' \bJ(x_1) \ldots \bJ(x_m) J(x'_1) \ldots J(x'_n) \langle 0 |  \chi(x'_1) \ldots \chi(x'_n) \bchi(x_1) \ldots \bchi(x_m)| 0 \rangle_{\qftn}.
\end{split}
\ee
One subtlety here is that, in the case of principal series, $\chi(x)$ and $\bchi(x)$ commute only up to a contact term. So we are careful to place $\bchi$ on the right and $\chi$ on the left --- a convention that we have already tacitly adopted above. Appendix \ref{appprincipalwavefunc} explains why this convention is superior to the opposite convention where the $\chi$ are placed on the right.

As usual, this functional can be related to a generating function of ``connected correlators'', $W$,  via
\be
Z[J, \bJ] = e^{W[J, \bJ]}.
\ee
The correlators we want can then be recovered via functional differentiation and after integrating with the functions that define the external states.

We normalize the vacuum so that
\be
Z[0] = 1 \qquad W[0] = 0,
\ee
which ensures that vacuum bubbles, which are disconnected from all other pieces of the diagram,  do not contribute to any correlator.

\section{Perturbative observables \label{secfeynman}}
In this section, we formulate Feynman rules to compute the norm of perturbative states and gauge-fixed perturbative cosmological correlators as defined in \eqref{ccexpr}. 

As we have explained above, these computations reduce to integrated QFT vacuum correlators. Vacuum correlators in a QFT can be computed using the in-in formalism \cite{Weinberg:2005vy} or by computing the wavefunction by analytic continuation from AdS \cite{Maldacena:2002vr}. Here, we will simply assume that the wavefunction for the Euclidean vacuum is given to us and we will remain agnostic about how it was computed.

The discussion of \cite{Chakraborty:2023yed,Chakraborty:2023los} focused on wavefunctions of both the metric and of matter fluctuations, $\chi$.  Since we are interested in the $G_N \to 0$ limit, we will ignore the metric fluctuations. This limit is slightly subtle: one must first square the wavefunction and then perform the trivial integral over the metric fluctuations to obtain a wavefunction that depends on the matter fields. Moreover, \cite{Chakraborty:2023yed,Chakraborty:2023los} carefully kept track of a leading phase-factor in the wavefunction. The correlators we study here are insensitive to the phase factor. So we will simply take the {\em effective} wavefunction of the matter fields to 
have the form
\be
\label{vacwaveform}
\Psi_{0}[\chi] = e^{\sum_{n=2}^{\infty} \int  d x_1 \ldots d x_n \lambda^{n-2} G_n(\vec{x}) \bchi(x_1) \ldots \bchi(x_n)},
\ee
where each $G_n$ obeys the Ward identities of a $n$-point conformal correlator of operators with dimension $\Delta$ \cite{Pimentel:2013gza,Baumann:2022jpr} and we have included a perturbative parameter $\lambda$ for book-keeping. 

A vacuum correlator is now computed via
\be
\label{vevviavacuum}
\langle 0 | \chi(x_1) \ldots \chi(x_n) \bchi(x'_1) \ldots \bchi(x'_m) | 0 \rangle  = \int [D \chi D \bchi] |\Psi_0[\chi]|^2 \chi(x_1) \ldots \chi(x_n) \bchi(x'_1) \ldots \bchi(x'_m).
\ee
As explained in Appendix \ref{appprincipalwavefunc}, this expression naturally computes correlators where the $\bchi$ appear to the right of the $\chi$. Second, the measure of integration must be normalized carefully and, when $\chi$ is in the principal series, gives rise to some contact-term contributions in the propagators that we describe below. Interactions can now be treated simply by expanding the non-Gaussian pieces in the wavefunction and the norm \eqref{normfixed} and \eqref{generalcosmo} can be computed through smeared and integrated linear combinations of such correlators.

\subsection{Feynman rules}
We present Feynman rules in position space. In some cases, it is convenient to perform the integral by Fourier transforming to momentum space and the momentum-space rules can be obtained easily from the rules given below. 

\paragraph{Propagators.}
For principal-value fields we need to keep track of the following kinds of propagators, each of which corresponds to a free-field two-point function. 

\[
\begin{array}{c|c|c}
	{\rm Free-field~expression} & {\rm Symbol} & {\rm Value} \\ \hline
	\langle 0 | \chi(x) \bchi(y) |0 \rangle & \includegraphics[width=0.2\textwidth]{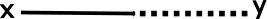} & \mathfrak{c}_{\x \bchi}\delta(x-y) \\
	\langle 0 | \chi(x) \chi(y) |0 \rangle & \includegraphics[width=0.2\textwidth]{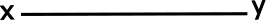} & {\mathfrak{c}_{\x\x} \over |x - y|^{2 \Delta}} \\
	\langle 0 | \bchi(x) \bchi(y) |0 \rangle & \includegraphics[width=0.2\textwidth]{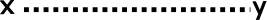} & {\mathfrak{c}_{\bchi\bchi} \over |x - y|^{2 \dbar}} \\
\end{array}
\]

The first line above involves a contact term that allows the contraction of $\chi$ and its conjugate \cite{Sun:2021thf}. It arises because if the wavefunction is written as a function of $\chi(x)$ then $\bchi(x)$ can be thought of as its canonical conjugate. The explicit forms of the constant coefficients $\mathfrak{c}_{\x \bchi},\mathfrak{c}_{\x\x}$ and $\mathfrak{c}_{\bchi\bchi}$ are given in  Section \ref{secsample}. As mentioned above,  $\bchi(x)$ always appears to the right of $\chi(x)$.

For the complementary series there is no conjugate field and no contact term. Therefore, only the $\chi-\chi$ propagator displayed above is relevant. 

\paragraph{Vertices.}
We also get interaction vertices by expanding the wavefunction in \eqref{vevviavacuum}. However, it is important to note that the bra and the ket themselves contribute as effective vertices in the computation of a norm or a correlator in \eqref{normfixed} or \eqref{generalcosmo}.  We distinguish the vertices that come from the bra, the ket, the wavefunction and its conjugate using different colors. This leads to the following Feynman rules.
\[
\begin{array}{c|c|c}
	{\rm  Origin} & {\rm Symbol} & {\rm Value} \\ \hline
	{\rm Ket} & \includegraphics[height=1cm]{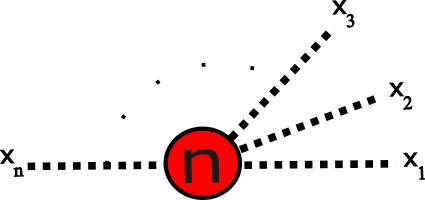} & \dcoeff[n](x_1,x_2,\ldots x_n) \\
	{\rm Bra}  & \includegraphics[height=1cm]{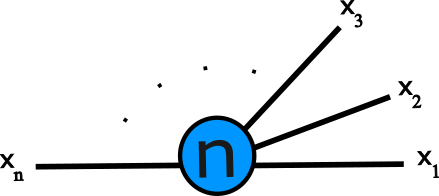} & \dcoeff[n]^*(x_1,x_2,\ldots x_n) \\
	\Psi[\chi] & \includegraphics[height=1cm]{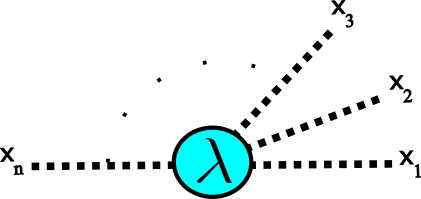} & \lambda^{n-2} G_n(x_1 \ldots x_n) \\
	\Psi^*[\chi] & \includegraphics[height=1cm]{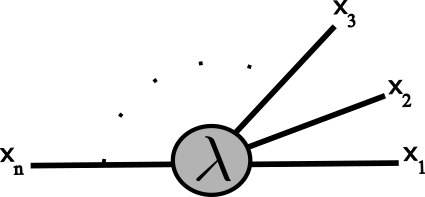} & \lambda^{n-2} G_n^*(x_1 \ldots x_n)
\end{array}
\]
The bra-ket and contributions from the wavefunction and its conjugate must all be integrated over the coordinates $x_i$. 

\paragraph{Fixed points.}
Not all points in the diagram are integrated. Each diagram must have at least three fixed points. In the computation of a norm, there are exactly three fixed points that arise from the formula \eqref{normfixed}. The computation of a cosmological correlator has three or more fixed points that arise from the insertions of the operators in \eqref{generalcosmo}. These fixed points will be displayed explicitly in each diagram.

The presence of these fixed points is important. Each vertex involves a conformally covariant function. Therefore, before we perform the position integrals, the entire integrand obtained by joining the vertices with propagators is conformally covariant. If all the external points were integrated over, we would get a divergence proportional to the volume of the conformal group. The presence of at least three-fixed points helps to avoid this divergence subject to some additional constraints that we discuss next.

As is standard, the contribution of a diagram is obtained by simply joining the vertices with propagators and performing the position integrals associated with each vertex. We provide examples in the next section.

\subsection{Group-volume divergences}
For some diagrams, the presence of three-fixed points might be insufficient to remove the divergence coming from the volume of the conformal group. This happens when the Feynman diagram breaks up into the product of two disconnected conformally covariant terms. In this section, we explain how to remove some simple group-volume divergences by restricting attention to states and observables that satisfy certain conditions. We divide the discussion into two parts: group-volume divergences in the norm and group-volume divergences in cosmological correlators. 
The Feynman diagrams discussed here follow from the rules given above. However, the reader might find some of them unfamiliar due to the presence of vertices associated with the bra and the ket. If so, it is possible to skip this subsection on a first reading, examine the examples in Section \ref{secsample} and then return to this subsection.

\subsubsection{Group-volume divergences in the norm}
Recall that the norm is computed by multiplying the QFT norm of a state by the factor $\normconst$ that is, itself, inversely proportional to the volume of the de Sitter group. Equivalently, the norm may be computed by fixing any three points in the conformally invariant integral that appears in its computation. 

\begin{figure}
	\begin{center}
		\includegraphics[width=0.3\textwidth]{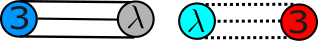}
		\caption{A diagram that displays a potential divergence due to the group volume. \label{bubblediagram}}
	\end{center}
\end{figure}

The diagram shown in Figure \ref{bubblediagram} provides an example of how disconnected diagrams might retain a divergence even after fixing three points.
This diagram appears in the computation of the norm of a three-particle state in the presence of an interaction that also contains a three-point term. The reader can consult the examples in the next section if this is not apparent. For now, the origin of the diagram is not crucial; it is only necessary to note that although three-points have been fixed in the contribution of the ket,  the diagram retains a divergence because the contraction of the bra and the interaction yields a conformally invariant integral and a corresponding group-volume divergence.

Group-volume divergences that arise from such disconnected diagrams can be avoided by restricting attention to those states that satisfy
\be
\label{stateorthvac}
\langle 0 | \psing \rangle_{\qftn} = 0,
\ee
and also
\be
\label{stateorthonep}
\langle 0| \chi(x) | \psing \rangle = \langle 0| \bchi(x) | \psing \rangle = 0.
\ee
Physically, these are very natural requirements to impose. The vacuum and one-particle states are non-normalizable. Therefore, it makes sense to subtract off their contribution from external states. Note that the overlap of a de-Sitter invariant state with a one-particle state can only be a constant, independent of $x$, since that is the only contribution allowed by conformal invariance \cite{ElShowk:2011ag}.

We now prove that imposing these conditions leads to the removal of disconnected diagrams. In terms of the generating function of connected correlators the computation of a norm corresponds to
\be
\begin{split}
&(\psing, \psing) = \normconst \sum_{m,m'} \int  d \vec{x} d \vec{x}' \,\dcoeff[m](\vec{x})  \dcoeff[m']^*(\vec{x}') \\ &\times \left. {\delta \over \delta J(x'_1)}  \ldots {\delta \over \delta J(x'_{m'})}  {\delta \over \delta \bJ(x_1)}  \ldots {\delta \over \delta \bJ(x_{m})}   e^{W[J, \bJ]} \right|_{J = \bJ = 0}.
\end{split}
\ee

We define
\be
{\delta \over \delta \bJ(x_1)} \ldots {\delta \over \delta \bJ(x_m)} e^{W[J, \bJ]} = \overline{P}[J, \bJ, x_1, \ldots x_m] e^{W[J, \bJ]},
\ee
where $\overline{P}[J, \bJ]$ is a functional of $J$ and $\bJ$ obtained by acting with $m$-functional derivatives on a polynomial of $W[J, \bJ]$. It is related to vacuum correlators through
\be
\overline{P}[J = 0, \bJ=0,x_1, \ldots x_m] = \langle 0 | \bchi(x_1) \ldots \bchi(x_m) | 0 \rangle.
\ee

If we had just been computing this vacuum correlator then those terms in $\overline{P}$ that contain multiple powers of $W$ would correspond to disconnected diagrams where some of the points $x_i$ are disconnected from each other.  However, in the computation under consideration all these diagrams are linked by the ket vertex. This is evident above in the presence of the smearing function $\dcoeff[m](\vec{x})$ that links all the external points. 

Now consider acting with one of the remaining-functional derivatives
\be
{\delta \over \delta J(x'_1)} \overline{P}[J, \bJ, x_1, \ldots x_n] e^{W[J]} = {\delta \overline{P}[J, \bJ, x_1, \ldots x_n] \over \delta J(x_1')} e^{W[J]} + \overline{P}[J, \bJ,  y_1 \ldots y_n] {\delta \over \delta J(x_1')} e^{W[J]}.
\ee
The first term on the right hand side comprises diagrams where the point $x_1'$ is connected to at least one of the $x_i$ points. But the functional $\dcoeff[m](\vec{x})$ connects the different $x_i$ insertions together and therefore the first term comprises diagrams where $x_1'$ is connected to the $x_i$. The second term generates a disconnected  diagram. 

Proceeding this way, we see that disconnected diagrams are only obtained if each ${\delta \over \delta {J(x_i')}}$ functional derivative acts on the $e^{W[J, \bJ]}$ factor. But, defining $P$ analogously to $\overline{P}$ above,  this implies that the set of disconnected diagrams factorizes into
\be
\begin{split}
&\sum_{n, m} \dcoeff[n]^*(\vec{x}') \dcoeff[m](\vec{x})  \overline{P}[J=0, \bJ=0, x_1, \ldots x_n] P[J=0, \bJ=0, x_1', \ldots x_m'] \\ &= \sum_{n} \dcoeff[n]^*(\vec{x}') P[J=0, \bJ=0, x_1', \ldots x_n'] \sum_{m} \dcoeff[m](\vec{x})  \overline{P}[J=0, \bJ=0, x_1, \ldots x_m],
\end{split}
\ee
but these terms correspond to the overlap of the state with the vacuum and its conjugate. Therefore, setting this to zero gets rid of diagrams where the bra and the ket are disconnected.

Even if the diagram is connected, a one-particle state is not normalizable since it is not possible to fix $3$ points in the integral for the norm. Condition \eqref{stateorthonep} disallows such states. Repeating the argument above, it can be seen that this condition also removes potential group-volume divergences from diagrams where the bra and the ket can be separated by cutting a single line.

\subsubsection{Group-volume divergences in cosmological correlators}
We now turn to cosmological correlators. The formal analysis in this case is similar to the analysis performed above for the norm and so we will simply state the results and provide some illustrative examples.

Some obvious group-volume divergences are removed by imposing
\be
\label{ccvacorth}
\langle 0 | \cprod | 0 \rangle_{\qftn} = 0.
\ee
In addition the smearing functions that enter \eqref{ccgeneral} must obey certain constraints. The integral of these smearing functions over any one coordinate must vanish 
\be
\label{cconepartorth}
\int d x_i g_q(\vec{x}) = 0 \qquad \forall i.
\ee
This must hold for each $g_q(\vec{x})$.  We emphasize that here the $d x_i$ is a single integral and so the condition above can be expanded as
\be
\int d x_1 g_q(\vec{x}) = \int d x_2 g_q(\vec{x}) \ldots = \int d x_q g_q(\vec{x}) = 0.
\ee
Second, for any pair of coordinates i,j we impose
\be
\label{cctwopartorth}
\int d x_i d x_j { g_q(\vec{x})\over |x_i - x_j|^{2 \Delta}} = 0.
\ee

The conditions \eqref{ccvacorth} may be satisfied by simply subtracting off a suitable multiple of the identity from $\cprod$. Equations \eqref{cconepartorth} and \eqref{cctwopartorth} can be satisfied by appropriate choices of smearing functions. 

Equations \eqref{cconepartorth} and \eqref{cctwopartorth} can also be expressed in Fourier space. Defining
\be
\widetilde{g}_q(\vec{k}) = \int g_q(\vec{x}) e^{i \vec{k} \cdot \vec{x}} \prod_j d x_j,
\ee
the conditions \eqref{cconepartorth} and \eqref{cctwopartorth} read
\be
\int d k_i \widetilde{g}_q(\vec{k}) \delta(k_i) = 0 \qquad \int d k_i d k_j \delta(k_i + k_j) \widetilde{g}_q(\vec{k}) |k_i - k_j|^{2 \Delta - d} = 0.
\ee
These conditions can be satisfied by choosing $\widetilde{g}(\vec{k})$ so that it vanishes whenever any of the $k_i$ vanish or when $k_i + k_j = 0$ for any pair of arguments:
\be
\label{ccsmearcondns}
\left. \widetilde{g}(\vec{k}) \right|_{k_i = 0} = 0; \qquad \left. \widetilde{g}(\vec{k}) \right|_{k_i + k_j= 0} = 0.
\ee

\begin{figure}[htbp]
	\centering
	\begin{subfigure}[b]{0.27\textwidth}
		\includegraphics[width=\textwidth]{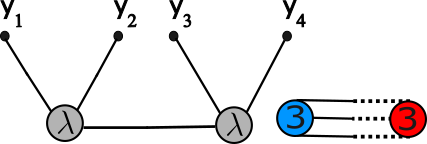}
		\caption{}
		\label{fignocontract}
	\end{subfigure}
	\hfill
	\begin{subfigure}[b]{0.27\textwidth}
		\includegraphics[width=\textwidth]{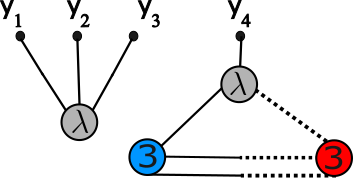}
		\caption{}
		\label{figonecontract}
	\end{subfigure}
	\hfill
	\begin{subfigure}[b]{0.27\textwidth}
		\includegraphics[width=\textwidth]{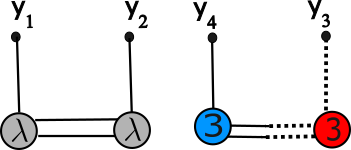}
		\caption{}
		\label{figtwocontract}
	\end{subfigure}
	
	\caption{The three kinds of diagrams that are removed by the conditions \eqref{ccvacorth}, \eqref{cconepartorth} and \eqref{cctwopartorth}. All these represent contributions to a four-point cosmological correlator in a three-particle state, where the vacuum has three-point interactions.}
	\label{fig:main}
\end{figure}

Repeating the arguments above now shows that some obvious group-volume divergences are removed by these conditions. Diagrams where the points in the cosmological correlator are completely disconnected from the bra and the ket vanish by virtue of \eqref{ccvacorth}. An example is shown in Figure \ref{fignocontract}.  Now consider those diagrams where only one or two external points from the cosmological correlator are connected to one part of the diagram whereas the other points are connected to another part; an example is shown in Figure \ref{figonecontract} and another example is shown in Figure \ref{figtwocontract}.  Such diagrams vanish by virtue of \eqref{cconepartorth} and \eqref{cctwopartorth}. This leaves us with diagrams that are either completely connected or those that break up into two or more disconnected pieces, each of which is connected to at least three points from the cosmological correlator. But these fixed points are enough to absorb the integral over the conformal group. 

\begin{figure}[H]
\centering
  \includegraphics[width=0.3\textwidth]{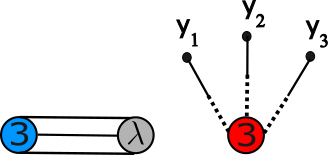}
\caption{A divergent contribution to a 3-point cosmological correlator in a 3-particle state that is removed by condition \eqref{stateorthvac}. \label{fig3ptdivergence}}
\end{figure}

Some other potential divergences in cosmological correlators, such as the one shown in Figure \ref{fig3ptdivergence} are removed by the condition \eqref{stateorthvac}. 

\paragraph{Additional divergences and renormalization.}
We emphasize that the removal of simple group-volume divergences does not remove all divergences associated with the norm, or with cosmological correlators. For instance, at higher-orders in perturbation theory, we see the usual UV divergences that appear in any effective field theory computation. In general, it is necessary to appropriately renormalize the wavefunction by adding counterterms to ensure that cosmological correlators are finite. We postpone an analysis of such effects to future work.

\section{Sample computations \label{secsample}}
In this section we provide some sample calculations that illustrate the use of the Feynman rules presented in the previous sections. We will compute norms and the values of cosmological correlators in a variety of states. 

A striking feature of these computations is that every simple state yields cosmological correlators that differ from QFT expectation values in the Euclidean vacuum. We present several examples where the Euclidean vacuum is Gaussian and therefore vacuum expectation values trivially factorize into products of two-point functions. Nevertheless, a nontrivial excited state in such a theory gives rise to nontrivial conformally invariant cosmological correlators. It is the conformal invariance that is special; even in a free nongravitational field theory in de Sitter space, excited states can give rise to nontrivial correlation functions for operator insertions on the late-time slice but these correlators are never conformally invariant since the underlying excited state cannot be conformally invariant while being normalizable.

For technical simplicity, we will study fields in the principal-series representation. In this case, we are able to borrow various tools and orthogonality relations from harmonic analysis that make it possible for us to compute explicit position space correlators. To this end we will use many of the results from \cite{Karateev:2018oml}. We will quote mathematical identities relevant to us in the main text itself but interested readers are encouraged to consult \cite{Karateev:2018oml} to understand them from a representation theory point of view.  Since these calculations are provided for the purpose of illustration, and since they are performed at lowest nontrivial order in perturbation theory, we set aside potential subtleties with principal-series fields that might arise at higher orders in perturbation theory \cite{SalehiVaziri:2024joi}.

This section is divided into two parts. In the first part, we study a Gaussian vacuum and then we study an interacting vacuum.  In both cases, we obtain beautiful closed-form answers for three and four point cosmological correlators in various states. The structural aspects of these answers deserve further investigation that we postpone to future work.

The norm of group-averaged states was previously studied from a Fock-space perspective in \cite{Marolf:2008hg}. 

\subsection{Cosmological correlators in excited  states about a Gaussian vacuum}
We start by considering the case of a Gaussian vacuum. We will consider the case where the vacuum has $n$-different principal-value fields, denoted by $\chi_i$, having conformal weight $\D_i={d\over2}+i\mu_i$. The nonzero correlators in the vacuum are \cite{Sun:2021thf} 
\be
\label{allprops}
\begin{split}
&\langle 0 | \chi_i(x) \chi_j(x')|0  \rangle = {\mathfrak{c}_{\x_i\x_i} \over |x - x'|^{2 \Delta_i}} \delta_{i j}~~~~~;\mathfrak{c}_{\x_i\x_i}={\G[\D_i]\G[{d\over2}-\D_i]\over 4\pi^{{d\over2}+1}} \\
&\langle 0 |\bchi_i(x)\bchi_j(x')|0  \rangle = {\mathfrak{c}_{\bchi_i\bchi_i} \over |x - x'|^{2 \dbar_i}} \delta_{i j}~~~~~;\mathfrak{c}_{\bchi_i\bchi_i}={\G[\D_i]\G[\D_i-{d\over2}]\over 4\pi^{{d\over2}+1}}\\
&\langle 0 | \chi_i(x)\bchi_j(x')|0  \rangle = \mathfrak{c}_{\x_i\bchi_i}\delta(x - x') \delta_{i j}~~; \mathfrak{c}_{\x_i\bchi_i}={\coth{\pi\mu_i}+1\over 4 \mu_i }.
\end{split}
\ee
All higher point correlators in the vacuum factorize into products of the two-point functions give above.  In terms of the wavefunction \eqref{vacwaveform}, this means that it has Gaussian terms but no non-Gaussian terms.

\subsubsection{3 particle state}
We will start with the simplest example of a nontrivial state. Consider a theory with a single field in the principal-series, $\chi(x)$ (and its conjugate field $\bchi$), having conformal weight $\D$ ($\dbar$ for the conjugate field). This theory has a single 3-particle state. We write it as
\be
|3\ra= \int  dx_1 dx_2 dx_3 \dcoeff[3](x_1,x_2,x_3) :\bchi(x_1)\bchi(x_2)\bchi(x_3): |0\ra.
\ee
Here $: :$ denotes normal ordering. The form of $\dcoeff[3](x_1,x_2,x_3)$ is fixed by conformal invariance to be 
\be
\label{eq:3particle_defn_coeff}
\dcoeff[3](x_1,x_2,x_3)= {\mcalc_{\D\D\D} \over \abs{x_1-x_2}^{\D}\abs{x_1-x_3}^{\D}\abs{x_2-x_3}^{\D}}.
\ee
The constant $\mcalc_{\D\D\D} $ fixes the overall normalization of the state.

\begin{figure}[H]
	\centering
	\includegraphics[scale=0.5]{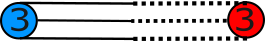}
	\caption{Diagram contributing the norm of the 3-particle state 
	\label{fig:3par_norm}}
\end{figure}

The bra corresponding to the ket above is 
\be
\langle 3 |  = \int  dx_1 dx_2 dx_3 \dcoeff[3]^*(x_1,x_2,x_3) \langle 0 | :\x(x_1)\x(x_2)\x(x_3):.
\ee
The norm $\langle 3|3\ra $ gets contribution from the diagram displayed in Figure \ref{fig:3par_norm}.
We find --- either by 
using the Feynman rules discussed in Section \ref{secfeynman} --- or by directly contracting the bra and the ket, 
\be
\label{eq:3particle_norm}
\begin{split}
	 \left(3, 3 \right) &=3! \mathfrak{c}_{\x\bchi}^3 \normconst \int{ dx_1 dx_2 dx_3}  |\dcoeff[3](x_1, x_2, x_3)|^2 
	\\&=  {3! \mathfrak{c}_{\x\bchi}^3\abs{\mcalc_{\D\D\D}}^2}.
\end{split}
\ee
Where to get the final equality, we gauge fixed the integral, which leaves behind only the constant coefficients. We refer the reader to section 5.3 of \cite{Chakraborty:2023yed} for details. 

Integrals over conformally invariant structures of the kind that appear above were also studied extensively in  \cite{Karateev:2018oml}. To make contact with their notation, we define
\be
\langle O^{\D_1}(x_1) O^{\D_2}(x_2) O^{\D_3}(x_3) \ra \equiv {C_{\D_1\D_2\D_3}\over \abs{x_1-x_2}^{\D_1+\D_2-\D_3}\abs{x_1-x_3}^{\D_1+\D_3-\D_2}\abs{x_2-x_3}^{\D_2+\D_3-\D_1}}.
\ee
The notation is designed to be reminiscent of the correlation function of CFT operators although we reiterate that there is no physical CFT here; it is simply that the gravitational constraints force the states to be smeared
with functions that obey the Ward identities of CFT correlators.

Using this notation, the result \eqref{eq:3particle_norm} can also be derived from the formula
\be
\label{eq:3point_pairing_integral}
\begin{split}
&\int { dx_1 dx_2 dx_3\over \dsvol} \langle O^{\D_1}(x_1) O^{\D_2}(x_2) O^{\D_3}(x_3)  \ra \langle  O^{\dbar_1}(x_1) O^{\dbar_2}(x_2) O^{\dbar_3}(x_3)   \ra
\\&={C_{\D_1\D_2\D_3}C_{\dbar_1\dbar_2\dbar_3} \over  \rotgrpvol}.
\end{split}
\ee

In our case, since we are considering principal-series fields $C_{\dbar \dbar \dbar} = C_{\D \D \D}^*$ which gives back \eqref{eq:3particle_norm}. 
One can choose the value of $\abs{C_{\D\D\D}}$ to normalize the state to have norm $1$, although we will keep it as it is.

\begin{figure}[H]
	\centering
	\includegraphics[scale=0.5]{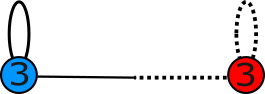}
	\caption{A potentially divergent contribution to the norm of the 3-particle state that is removed by normal ordering	\label{fig:3_3_self_contrn}}
\end{figure}

Normal ordering the operators is important to ensure that additional diagrams, with self-contractions, such as Figure \ref{fig:3_3_self_contrn}
do not contribute to the norm.  Normal ordering also automatically ensures that the conditions of Section \ref{secfeynman} are met for this state.

\subsubsection{Cosmological correlators in the 3-particle state}
We now turn to the computation of cosmological correlators. The simplest one to compute is a three-point cosmological correlator. However, we find that
\be
\llangle 3 | \chi(y_1) \chi(y_2) \chi(y_3) | 3 \rrangle = 0.
\ee
This can be seen by expanding the expression above, which yields the expectation value of 9 fields in the vacuum. This expectation value vanishes since the number of insertions is odd (Recall that the vacuum is free).

So we now turn to the computation of a four-point cosmological correlator 
\[
\llangle 3| :\x(y_1)\x(y_2)\x(y_3) \x(y_4): |3\rrangle.
\]
Through normal ordering, we again ensure that the operator has no vacuum expectation value. In addition, it is to be understood that the result of this correlator is always smeared with a function of $y_i$ that obeys the conditions \eqref{ccsmearcondns}.

\begin{figure}[H]
	\centering
	\includegraphics[scale=0.5]{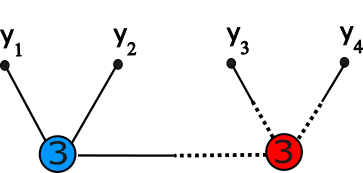}
	\caption{ Diagram contributing to a 4-point function in a 3-particle state \label{fig:3par_4point_fn}}
\end{figure}

One of the Feynman diagrams that contributes to this correlator is displayed in Figure \ref{fig:3par_4point_fn}. Denoting its value by $\mathcal{F}_1$, we find that

\be
\begin{split}
	\mathcal{F}_1&
	= (3!)^2\int \left(\prod_{i=1}^3  dx_i  dx'_i\right)
	\g G_3^*(x_1,x_2,x_3) \g G_3(x'_1,x'_2,x'_3)  
	\\& \times ~ \wick{\c1 \x(y_1) \c1 \x(x_1)} \wick{\c1 \x(y_2) \c1 \x(x_2) } \wick{\c1 \x(x_3) \c1 \bchi(x'_3) } \wick{\c1 \x(y_3)\c1 \bchi(x'_1) } \wick{\c1 \x(y_4)\c1 \bchi(x'_2) } 
	\\&=(3!)^2 \mathfrak{c}_{\x\x}^2\mathfrak{c}_{\x\bchi}^3\int \left(\prod_{i=1}^3  dx_i  dx'_i\right)
	\g G_3^*(x_1,x_2,x_3) \g G_3(x'_1,x'_2,x'_3)  
	\\& \times ~ {1\over \abs{y_1-x_1}^{2\D}} {1\over \abs{y_2-x_2}^{2\D}} \g^d(x_3-x'_3) \g^d(y_3-x'_1) \g^d(y_4-x'_2)
	\\&=(3!)^2 \mathfrak{c}_{\x\x}^2\mathfrak{c}_{\x\bchi}^3\int \left(\prod_{i=1}^3  dx_i \right)\g G_3^*(x_1,x_2,x_3) \g G_3(y_3,y_4,x_3) {1\over \abs{y_1-x_1}^{2\D}} {1\over \abs{y_2-x_2}^{2\D}}.
\end{split}
\ee

The three remaining integrals are known conformal integrals. First, we have integrals that, in the CFT literature, are called ``shadow transforms''. This converts a three-point function of operators with dimension $\D_1, \D_2, \D_3$ to a three-point function of operators with dimensions $\dbar_1, \D_2, \D_3$.  We write this as
\be
\begin{split}
&\langle O^{S[\D_1]}(y_1) O^{\D_2}(x_2) O^{\D_3}(x_3) \rangle \equiv \int d x_1 \langle O^{\D_1} (x_1)  O^{\D_2}(x_2) O^{\D_3} (x_3) \rangle {1 \over \abs{y_1 - x_1}^{2 \dbar_1}} \\
&= {C_{\D_1 \D_2 \D_3} N(\D_1 + \D_2 - \D_3, \D_1 + \D_3 - \D_2, 2 \dbar_1) \over \abs{y_1 - x_2}^{\dbar_1 + \D_2 - \D_3} \abs{y_1 - x_3}^{\dbar_1 + \D_3 - \D_2} \abs{x_2 - x_3}^{\D_2 + \D_3 - \dbar_1}} \\
&\equiv {C_{S[\D_1] \D_2 \D_3}  \over \abs{y_1 - x_2}^{\dbar_1 + \D_2 - \D_3} \abs{y_1 - x_3}^{\dbar_1 + \D_3 - \D_2} \abs{x_2 - x_3}^{\D_2 + \D_3 - \dbar_1}},
\end{split}
\ee
where
\be
N(a,b,c)={\pi^{d\over2} \G[{d-a\over2}]\G[{d-b\over2}] \G[{d-c\over2}] \over \G[{a\over2}]\G[{b\over2}]\G[{c\over2}]}.
\ee
The notation $S[\D]$ keeps track of the fact that  shadow transformations have been performed and the resulting numerical factors are absorbed into the three-point function. Multiple shadow transforms can be performed in sequence. We will use the same notation below to indicate the shadow transformation of more-general expressions.

In the case at hand we have two shadow transforms and obtain
\be
\begin{split}
	\g G^*(y_1^{S[\D]},y_2^{S[\D]},x_3)& \equiv \int  dx_1  dx_2 \delta G^*_3(x_1, x_2, x_3) {1\over \abs{y_1-x_1}^{2\D}} {1\over \abs{y_2-x_2}^{2\D}}
	\\&={\mcalc^*_{\D\D\D}N(\dbar,\dbar,2\D) N(\D,2\D,2\dbar-\D)\over \abs{x_3-y_1}^{\dbar} \abs{x_3-y_2}^{\dbar} \abs{y_1-y_2}^{2\D-\dbar}}
	\\&\equiv{\mcalc^*_{S[\D]S[\D]\D}\over \abs{x_3-y_1}^{\dbar} \abs{x_3-y_2}^{\dbar} \abs{y_1-y_2}^{2\D-\dbar} }.
\end{split}
\ee

Thus we arrive at the expression,
\be
	\mathcal{F}_1=(3!)^2 \mathfrak{c}_{\x\x}^2\mathfrak{c}_{\x\bchi}^3\int  dx_3\g G_3^*(y_1^{S[\D]},y_2^{S[\D]},x_3) \g G_3(x_3,y_3,y_4).
\ee

The last integral is an integral of the product of two three-point functions over a common point. This  gives rise to a four-point partial wave using the formula \cite{Karateev:2018oml} 
\be
\begin{split}
\label{eq:CPW_defn}
\fourpw^{\Delta_1\D_2\D_3\D_4}_{\Delta}(x_1,x_2,x_3,x_4) = \int d x \langle O^{\Delta_1}(x_1) O^{\Delta_2}(x_2) O^{\Delta}(x) \rangle  \langle O^{\Delta_3}(x_3) O^{\Delta_4}(x_4) O^{\dbar}(x) \rangle.
\end{split}
\ee
Hence,
\be
\begin{split}
	\mathcal{F}_1&=(3!)^2 \mathfrak{c}_{\x\x}^2\mathfrak{c}_{\x\bchi}^3\int  dx_3\g G_3^*(y_1^{S[\D]},y_2^{S[\D]},x_3) \g G_3(x_3,y_3,y_4) 
	\\&=\mathcal{K}_1 \fourpw_{\dbar}^{\Delta \Delta \Delta \Delta}(y_1, y_2, y_3, y_4),
\end{split}
\ee
where,
\be
\mathcal{K}_1 = (3!)^2 \mathfrak{c}_{\x\x}^2\mathfrak{c}_{\x\bchi}^3{\mcalc^*_{S[\D]S[\D]\D} \over C_{\D\D\dbar} }.
\ee

The full $4$-point cosmological correlator can be easily written down taking into account other Feynman diagrams obtained by simply permuting the external legs.
\be
\begin{split}
&\llangle 3|:\x(y_1)\x(y_2)\x(y_3) \x(y_4): |3 \rrangle
\\&=\mathcal{K}_1 \left(\fourpw_{\dbar}^{\D\D\D\D}(y_1,y_2,y_3,y_4)+\fourpw_{\dbar}^{\D\D\D\D}(y_1,y_3,y_2,y_4)+\fourpw_{\dbar}^{\D\D\D\D}(y_1,y_4,y_2,y_3)\right).
\end{split}
\ee
Note that the normal ordering prescription automatically removes the term in the four-point function that could have been the product of two two-point functions.

Explicit formulas for the four-point partial waves in terms of hypergeometric functions for $d=2,4,6$ can be found in \cite{Dolan:2003hv}.

So we see that the three-particle state gives rise to a beautiful and explicitly conformally invariant formula for the four-point cosmological correlator.

\subsubsection{4 particle state}
For our next example, it is convenient to consider a theory with two scalar fields $\x_1$ and $\x_2$, both in the principal-series. The reason for studying two fields rather than one is simply that it allows us to perform the relevant conformal integrals analytically as we explain at the end of the computation.

A four particle state in the theory is,
\be
|4\ra=\int \left(\prod_{i=1}^4  dx_i\right) \g G_4(x_1^{\D_1},x_2^{\D_1},x_3^{\D_2},x_4^{\D_2}) :\bchi_1(x_1)\bchi_1(x_2)\bchi_2(x_3)\bchi_2(x_4): |0\ra,
\ee
where, for clarify, we have placed a superscript on the points to indicate the conformal weights of the function $G_4$. 
Conformal invariance restricts the form of $ \g G_4(x_1^{\D_1},x_2^{\D_1},x_3^{\D_2},x_4^{\D_2})$ up to arbitrary functions of the cross ratios. One convenient choice is to parameterize $\g G_4$ in terms of the conformal partial waves.
\be
\g G_4(x_1^{\D_1},x_2^{\D_1},x_3^{\D_2},x_4^{\D_2}) =\int_{d\over2}^{{d\over2}+i\infty} {d\D\over 2\pi i} I(\D) \fourpw_{\D}^{\D_1\D_1\D_2\D_2}\left(x_1,x_2,x_3,x_4\right).
\ee
Where $I(\D)$ is a function that parameterizes the state. We are allowed to choose $I(\D)$ and different choices define different states.  Since we defined the partial wave using \eqref{eq:CPW_defn}, the normalization of $\fourpw_{\D}^{\D_1\D_1\D_2\D_2}$ also depends on the choice of some coefficients $C_{\D_1\D_1\D}$ and $C_{\dbar \D_2 \D_2 }$. These can always be reabsorbed into a redefinition of $I(\D)$ since the only combination that can appear is the product $I(\D) C_{\D_1\D_1\D} C_{\dbar \D_2\D_2}$.

Harmonic analysis on $\confgrp$ tells us that the conformal partial waves satisfy an orthogonality relation \cite{Karateev:2018oml} (We denote $\D={d\over2}+is$ and $\dbar'={d\over2}-is'$).
\be
\label{eq:orthogonality_CPW}
\begin{split}
	&\int {\prod_{i=1}^4 d x_i\over \dsvol} \fourpw_{\D}^{\D_1\D_2\D_3\D_4}(x_1,x_2,x_3,x_4)
	\fourpw_{\dbar'}^{\dbar_1\dbar_2\dbar_3\dbar_4}(x_1,x_2,x_3,x_4)
	\\&={C_{\D_1\D_2\D}C_{\dbar_1\dbar_2\dbar}\over  \rotgrpvol} {C_{\D_3\D_4\dbar}C_{\dbar_3\dbar_4\D}\over  \rotgrpvol} {2\pi \g(s-s')\over \mu(\D)}.
\end{split}
\ee
Where $\mu(\D)$ --- the Plancherel measure of the representation --- is given by
\be
\label{eq:plancherel_measure}
\mu(\D)={\G[\D]\G[\dbar]\over \pi^d \G[{d-2\D\over2}]\G[{d-2\dbar\over2}]\text{vol$\left(\text{SO}(\text{d})\right)$}}.
\ee

\begin{figure}[H]
	\centering
	\includegraphics[scale=0.5]{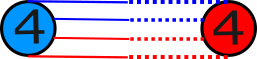}
	\caption{Norm of the 4-particle state}
	\label{fig:4parnorm}
\end{figure}

The norm of the state can now be easily computed using the orthogonality relation of the conformal partial waves.
The relevant diagram is shown in Figure \ref{fig:4parnorm}.  (We use blue lines for $\x_1$ and red lines for $\x_2$)
Applying the Feynman rules on this diagram, we find
\be
\begin{split}
\left(4,4\right)&=(2!)^2 \mathfrak{c}_{\x_1\bchi_1}^2 \mathfrak{c}_{\x_2\bchi_2}^2 \mathcal{K}\int \left(\prod_{i=1}^4 dx_i\right) \g G_4^*(x_1^{\D_1},x_2^{\D_1},x_3^{\D_2},x_4^{\D_2})
\g G_4(x_1^{\D_1},x_2^{\D_1},x_3^{\D_2},x_4^{\D_2})
\\&=(2!)^2 \mathfrak{c}_{\x_1\bchi_1}^2 \mathfrak{c}_{\x_2\bchi_2}^2 \rotgrpvol\int_{d\over2}^{{d\over2}+i\infty} {d\D\over 2\pi i} I(\D)^*
\int_{d\over2}^{{d\over2}+i\infty} {d\D'\over 2\pi i} I(\D')
\\&\times
 \int { dx_1 dx_2 dx_3 dx_4\over \dsvol} \left(\fourpw_{\D}^{\D_1\D_1\D_2\D_2}(x_1,x_2,x_3,x_4)\right)^* \times\left(\fourpw_{\D'}^{\D_1\D_1\D_2\D_2}(x_1,x_2,x_3,x_4)\right)
\\&=(2!)^2 {\mathfrak{c}_{\x_1\bchi_1}^2 \mathfrak{c}_{\x_2\bchi_2}^2\over \rotgrpvol}\int_{d\over2}^{{d\over2}+i\infty} {d\D\over 2\pi i} {\abs{I(\D)}^2\over \mu(\D)} \abs{C_{\D_1\D_1\D}}^2\abs{C_{\dbar\D_2\D_2}}^2.
\end{split}
\ee
Here, we have used the orthogonality relation of the conformal partial waves in the final step.
As we can see from the final expression, one can appropriately choose $I(\D) C_{\D_1\D_1\D} C_{\dbar\D_2\D_2}$ so as to get a finite norm for the state. 

The Feynman diagrams for a single-field model are similar to the two-field model.  However, these diagrams would lead to the integral of the product of two partial waves where the points on the left are paired with the points on the right in all possible combinations.  But the partial wave is only symmetric under the interchange of points $1 \leftrightarrow 2$ and $3 \leftrightarrow 4$.  Therefore, while such an integral can still be evaluated numerically, one cannot obtain a simple analytic answer using the orthogonality relation above. This explains our choice of the two-field model for illustrative purposes. The same considerations guide us in the examples below.

\subsubsection{Cosmological correlators in the 4 particle state}
The three point cosmological correlator vanishes in the 4 particle state when the vacuum is Gaussian.
\be
\llangle 4|  \x_1(y_1)\x_1(y_2)\x_1(y_3) | 4 \rrangle = 0.
\ee
Expanding the expression above leads to a vacuum correlator with 11 insertions which vanishes since it cannot be factorized into a product of 2-point functions.

\begin{figure}[h]
	\centering
	\includegraphics[scale=0.5]{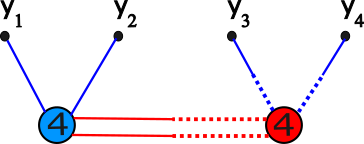}
	\caption{Four-point cosmological correlator in a four-particle state \label{fig:4par4pointfn}}
\end{figure}

So, we turn to the four point function $\llangle 4| :\x_1(y_1)\x_1(y_2)\x_1(y_3)\x_1(y_4): | 4 \rrangle$. The relevant Feynman diagram for the $4$-point function is shown in Figure \ref{fig:4par4pointfn}. 
Denoting its value by $\mathcal{F}_2$, we see that
\be
	\mathcal{F}_2 =(2!)^3 \mathfrak{c}_{\x_1\bchi_1}^2 \mathfrak{c}_{\x_2\bchi_2}^2 \mathfrak{c}_{\x_1\x_1}^2\int  dx_1 dx_2 \g G_4^*(y_1^{S[\D_1]},y_2^{S[\D_1]},x_1^{\D_2},x_2^{\D_2}) \g G_4(y_3^{\D_1},y_4^{\D_1},x_1^{\D_2},x_2^{\D_2}),
\ee
where we have used $S[\D_1], S[\D_2]$ to indicate that $\g G_4$ has been shadow transformed.
Parsing this further, we find
\be
\label{eq:4pointfn_4particle_feyn}
\begin{split}
	\mathcal{F}_2&=(2!)^3 \mathfrak{c}_{\x_1\bchi_1}^2 \mathfrak{c}_{\x_2\bchi_2}^2 \mathfrak{c}_{\x_1\x_1}^2\int_{d\over2}^{{d\over2}+i\infty} {d\D\over 2\pi i} I(\D)^*  \int_{d\over2}^{{d\over2}+i\infty} {d\D'\over 2\pi i} I(\D') 
	\\&\int  dx_1 dx_2 \left( \left(\fourpw_{\D}^{S[\D_1]S[\D_1]\D_2\D_2}(y_1,y_2,x_1,x_2)\right)^* \fourpw_{\D'}^{\D_1\D_1\D_2\D_2}(y_3,y_4,x_1,x_2)  \right)
	\\&=(2!)^3 \mathfrak{c}_{\x_1\bchi_1}^2 \mathfrak{c}_{\x_2\bchi_2}^2 \mathfrak{c}_{\x_1\x_1}^2 \int_{d\over2}^{{d\over2}+i\infty} {d\D\over 2\pi i}
	\abs{I(\D)}^2 {\abs{C_{\dbar \D_2\D_2}}^2 \over \rotgrpvol \mu(\D)} 
\\ &\times \int d x \langle O^{S[\D_1]} (y_1) O^{S[\D_1]}(y_2) O^{\D}(x) \rangle^* \langle O^{\D}(x) O^{\D_1}(y_3) O^{\D_1}(y_4) \rangle
	\\&= \int_{d\over2}^{{d\over2}+i\infty} {d\D\over 2\pi i}
	\abs{I(\D)}^2 {\abs{C_{\dbar \D_2\D_2}}^2 \over \rotgrpvol \mu(\D)} \mathcal{K}_2(\D) \fourpw_{\dbar}^{\D_1\D_1\D_1\D_1}(y_1,y_2,y_3,y_4).
\end{split}
\ee
Here,
\be
\mathcal{K}_2(\D)=(2!)^3 \mathfrak{c}_{\x_1\bchi_1}^2 \mathfrak{c}_{\x_2\bchi_2}^2 \mathfrak{c}_{\x_1\x_1}^2 {C^*_{S[\D_1]S[\D_1]\D} \over C_{\D_1\D_1\dbar}}
\ee
Here, we first write the conformal partial wave as an integral over the $3$ point functions as given in \eqref{eq:CPW_defn}. Then, we use the following result for integrating a product of two three point structures, which is called a ``bubble integral'' \cite{Karateev:2018oml}.
	\be
	\label{eq:bubble integral formula}
	\begin{split}
		&\int  dx_1 dx_2 \langle O^{\D_1}(x_1)O^{\D_2}(x_2)O^{\D}(x)\ra \langle O^{\dbar_1}(x_1) O^{\dbar_2}(x_2) O^{\dbar'}(x') \ra
		\\&={C_{\D_1\D_2\D}C_{\dbar_1\dbar_2\dbar}\over \rotgrpvol \mu(\D)} \g^d(x-x') 2\pi \g(s-s').
\end{split}
	\ee
The factor $\mathcal{K}_2(\Delta)$ has various $3$-point function coefficients including those arising from shadow transforms. It ensures the proper normalisation of the conformal partial wave as given in \eqref{eq:CPW_defn}. Also note that, once we account for the normalization of $\fourpw_{\dbar}^{\D_1\D_1\D_1\D_1}$, it is only the product $I(\D) C_{\D_1\D_1\D} C_{\dbar \D_2\D_2}$ that enters the expression as anticipated.

One can now write down $4$-point correlator, taking into account other diagrams obtained by permuting the external lines. 
\be
\begin{split}
	&\llangle 4|:\x_1(y_1)\x_1(y_2)\x_1(y_3) \x_1(y_4): |4 \rrangle
	\\&=\int_{d\over2}^{{d\over2}+i\infty} {d\D\over 2\pi i} I_4(\D)
	\left(\fourpw_{\dbar}^{\D_1\D_1\D_1\D_1}(y_1,y_2,y_3,y_4)+y_2\leftrightarrow y_3+y_2\leftrightarrow y_4\right),
\end{split}
\ee
where 
\be
I_4(\D)= \abs{I(\D)}^2 {\abs{C_{\dbar\D_2\D_2}}^2\over  \rotgrpvol \mu(\D) } \mathcal{K}_2(\Delta).
\ee
So the four-point cosmological correlator is also given in terms of conformal partial waves, with the coefficient function $I_4(\D)$ being determined in terms of the parameters defining the state.

\subsubsection{$3+4$ particle state}
In the previous sections, we computed four point cosmological correlator in simple 3 particle state and 4 particle state. Both these states have a vanishing three point correlator as the vacuum is Gaussian.  It is possible to obtain a non-zero three-point function in a state that is a linear combination of  $3$ and $4$ particle states.
\be
\begin{split}
    &|\psi\ra= a|3\ra+b |4\ra
	\\& |3\ra=\int \left(\prod_{i=1}^3  dx_i\right) \g G_3(x_1^{\D_1},x_2^{\D_2},x_3^{\D_2}) :\bchi_1(x_1) \bchi_2(x_2) \bchi_2(x_3): |0\ra
	\\& |4\ra=\int \left(\prod_{i=1}^4  dx_i\right) \g G_4(x_1^{\D_1},x_2^{\D_1}x_3^{\D_2},x_4^{\D_2}) :\bchi_1(x_1) \bchi_1(x_2) \bchi_2(x_3) \bchi_2(x_4): |0\ra.
\end{split}
\ee
Here we have taken a two-field model and defined states with some foresight so as to obtain an analytic answer. The form of the three-point function is fixed by conformal invariance and its coefficient can be computed  numerically in more general states, including a single-field model.  

As in the previous two examples,  $\g G_3$ and $\g G_4$ can be written as
\be
\begin{split}
		&\g G_3(x_1^{\D_1},x_2^{\D_2},x_3^{\D_2})
		= \langle O^{\D_1} (x_1) O^{\D_2} (x_2) O^{\D_2}(x_3) \rangle =  {\mcalc_{\D_1\D_2\D_2}\over \abs{x_1-x_2}^{\D_1} \abs{x_1-x_3}^{\D_1} \abs{x_2-x_3}^{2\D_2-\D_1} }
		\\& \g G_4(x_1^{\D_1},x_2^{\D_1}x_3^{\D_2},x_4^{\D_2})
		=\int_{d\over2}^{{d\over2}+i\infty} {d\D\over 2\pi i} I(\D) \fourpw_{\D}^{\D_1\D_1\D_2\D_2}(x_1,x_2,x_3,x_4).
\end{split}
\ee

We can first compute the norm of the state $|\psi\ra$. Since the vacuum is Gaussian, there are no cross terms of the form $\left(3,4\right)$, and we get
\be
\left(\psi,\psi\right)=\abs{a}^2 \left(3,3\right)+\abs{b}^2\left(4,4\right).
\ee
The computation of $\langle 3|3\ra$ and $\langle 4|4\ra$ follows in the exact same manner as we have done in the previous sections. We will quote the answer below. 

\be
\begin{split}
\left(\psi,\psi\right)= &2! \abs{a}^2 { \mathfrak{c}_{\chi_2\bar{\chi}_2}^2 \mathfrak{c}_{\chi_1\bar{\chi}_1} \abs{\mcalc_{\D_1\D_2\D_2}}^2} \\ &+\abs{b}^2 \left((2!)^2 {\mathfrak{c}_{\x_1\bchi_1}^2 \mathfrak{c}_{\x_2\bchi_2}^2\over \rotgrpvol}\int_{d\over2}^{{d\over2}+i\infty} {d\D\over 2\pi i} {\abs{I(\D)}^2\over \mu(\D)} \abs{C_{\D_1\D_1\D}}^2\abs{C_{\dbar\D_2\D_2}}^2 \right).
\end{split}
\ee

We now move on to the computation of the $3$-point function.  This will get contributions only from the cross terms. More explicitly,
\be
\llangle\psi|\x(y_1)\x(y_2)\x(y_3)|\psi\rrangle=a^*b \llangle3|\x(y_1)\x(y_2)\x(y_3)|4\rrangle+b^*a \llangle4|\x(y_1)\x(y_2)\x(y_3)|3\rrangle.
\ee
Second, conformal invariance forces
\be
\llangle\psi|\x(y_1)\x(y_2)\x(y_3)|\psi\rrangle = {c\over \abs{y_1-y_2}^{\D}\abs{y_1-y_3}^{\D}\abs{y_2-y_3}^{\D}}.
\ee
Our task is to determine $c$ in terms of the parameters that define the state. As a check we see that only the combinations $a \mcalc_{\D_1\D_2\D_2}$ and $I(\D)C_{\D_1\D_1\D}C_{\dbar\D_2\D_2}$ can appear in the answer.

\begin{figure}[h]
	\centering
	\includegraphics[scale=0.5]{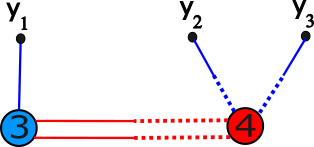}
	\caption{A contribution to the 3-point function in the 3+4 particle state}
	\label{fig:34par3pointfn1}
\end{figure}

The diagram that captures the term  $\llangle3|\x_1(y_1)\x_1(y_2)\x_1(y_3)|4\rrangle$  is shown in Figure \ref{fig:34par3pointfn1}. Denoting its value by $\mathcal{F}_3$ we have,
\be
\begin{split}
	\mathcal{F}_3&=(2!)^2\mathfrak{c}_{\x_1\bchi_1}^2 \mathfrak{c}_{\x_2\bchi_2}^2  \mathfrak{c}_{\x_1\x_1}\int  dx_1 dx_2 \g G_3^*(y_1^{S[\D_1]},x_1^{\D_2},x_2^{\D_2}) \g G_4(y_2^{\D_1},y_3^{\D_1},x_1^{\D_2},x_2^{\D_2})
	\\&= (2!)^2\mathfrak{c}_{\x_1\bchi_1}^2 \mathfrak{c}_{\x_2\bchi_2}^2  \mathfrak{c}_{\x_1\x_1}\int_{d\over2}^{{d\over2}+i\infty} {d\D\over 2\pi i} I(\D) \\&\times \left(\int  dx_1  dx_2 \langle O^{S[\D_1]}(y_1) O^{\D_2}(x_1)O^{\D_2}(x_2) \rangle^* \fourpw_{\D}^{\D_1\D_1\D_2\D_2} (y_2,y_3,x_1,x_2) \right) 
	\\&={\mathcal{K}_3\over \abs{y_1-y_2}^{\D_1} \abs{y_1-y_3}^{\D_1} \abs{y_2-y_3}^{\D_1} }.
\end{split}
\ee
Here, we have used the bubble integral formula \eqref{eq:bubble integral formula} to integrate over $x_1$ and $x_2$ and arrive at the 3rd equality. In the final step, the numerical factor is given by 
\be
\mathcal{K}_3={ (2!)^2\mathfrak{c}_{\x_1\bchi_1}^2 \mathfrak{c}_{\x_2\bchi_2}^2  \mathfrak{c}_{\x_1\x_1} I(\D_1) \mcalc^*_{S[\D_1]\D_2\D_2} C_{\dbar_1\D_2\D_2} C_{\D_1\D_1\D_1} \over \rotgrpvol \mu(\D_1)   }.
\ee

\begin{figure}[h]
	\centering
	\includegraphics[scale=0.5]{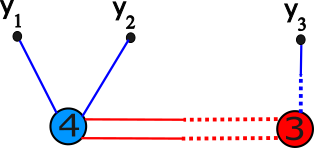}
	\caption{Another contribution to the 3-point function in the 3+4 particle state \label{fig34par3pointfn2}}
\end{figure}

The diagram that captures the term $\llangle4|\x_1(y_1)\x_1(y_2)\x_1(y_3)|3\rrangle$ is shown in Figure \ref{fig34par3pointfn2}. 
Denoting its value by $\mathcal{F}_4$, we have   
\be
\begin{split}
	F_4&=(2!)^2 \mathfrak{c}_{\x_1\x_1}^2 \mathfrak{c}_{\x_2\bchi_2}^2 \mathfrak{c}_{\x_1\bchi_1} \int  dx_1  dx_2 \g G_4^*(y_1^{S[\D_1]},y_2^{S[\D_1]},x_1^{\D_2},x_2^{\D_2}) \g G_3(y_3^{\D_1},x_1^{\D_2},x_2^{\D_2})
	\\&=(2!)^2 \mathfrak{c}_{\x_1\x_1}^2 \mathfrak{c}_{\x_2\bchi_2}^2 \mathfrak{c}_{\x_1\bchi_1}\int_{d\over2}^{{d\over2}+i\infty} {d\D\over 2\pi i} (I(\D))^*
	\\&\times \left(\int  dx_1 dx_2 \left(\fourpw_{\D}^{S[\D_1]S[\D_1]\D_2\D_2}(y_1,y_2,x_1,x_2)\right)^* \g G_3(y_3^{\D_1},x_1^{\D_2},x_2^{\D_2})  \right)
	\\&={\mathcal{K}_4\over \abs{y_1-y_2}^{\D_1}\abs{y_1-y_3}^{\D_1}\abs{y_2-y_3}^{\D_1}}.
\end{split}
\ee
Here,
\be
\mathcal{K}_4={ (2!)^2 \mathfrak{c}_{\x_1\x_1}^2 \mathfrak{c}_{\x_2\bchi_2}^2 \mathfrak{c}_{\x_1\bchi_1} I(\dbar_1)^* |C_{\D_1\D_2\D_2}|^2 C^*_{S[\D_1]S[\D_1]\dbar_1}	\over 
	\rotgrpvol \mu(\D_1)}.
\ee
Adding these two contributions and taking into account other diagrams which are related to these by permutation of the external legs we get,
\be
\begin{split}
	& \llangle \psi| \x_1(y_1)\x_1(y_2)\x_1(y_3) |\psi \rrangle={c\over \abs{y_1-y_2}^{\D_1}\abs{y_1-y_3}^{\D_1}\abs{y_2-y_3}^{\D_1}}
	\\& c= 3\times \left({a^*b \mathcal{K}_3+b^*a\mathcal{K}_4}\right).
\end{split}
\ee
Here again we see that $c$ is given purely in terms of the data defining the state. This example suggest that even if the vacuum has a vanishing $3$-point function, an appropriately chosen 'excited' state can have large non-Gaussianity.
\subsection{Cosmological correlators in the presence of a non-Gaussian vacuum}
In the previous examples we took the vacuum to be Gaussian. In this subsection, we study a case where the vacuum wavefunction has an interaction term and the leading contribution to the correlator comes from this interaction.
	
We now introduce a third field $\x_3$ which is again motivated by our desire to obtain explicit answers. Consider a state $|\psi\ra$ defined as follows. 
\be
\label{nongaussiansetup}
\begin{split}
	&|\psi\ra=a|3\ra+b|4\ra
	\\&|3\ra=\int  dx_1 dx_2 dx_3 \g G_3(x^{\D_1}_1,x_2^{\D_2},x_3^{\D_2}): \bchi_1(x_1)\bchi_2(x_2)\bchi_2(x_3):|0\ra
	\\&|4\ra=\int  dx_1 dx_2 dx_3 dx_4 \g G_4(x^{\D_1}_1,x_2^{\D_1},x_3^{\D_3},x_4^{\D_3}) :\bchi_1(x_1)\bchi_1(x_2)\bchi_3(x_3)\bchi_3(x_4):|0\ra
	\\& \g G_3(x_1^{\D_1},x_2^{\D_2},x_3^{\D_2})=
	{\mcalc_{\D_1\D_2\D_2}\over \abs{x_1-x_2}^{\D_1} \abs{x_1-x_3}^{\D_1} \abs{x_2-x_3}^{2\D_2-\D_1} }
	\\& \g G_4(x_1^{\D_1},x_2^{\D_1},x_3^{\D_3},x_4^{\D_3})
	=\int_{d\over2}^{{d\over2}+i\infty} {d\D\over 2\pi i} I(\D) \fourpw_{\D}^{\D_1\D_1\D_3\D_3}(x_1,x_2,x_3,x_4).
\end{split}
\ee
We take the vacuum to have a four point interaction term as given below. In the notation of \eqref{vacwaveform} this means that
\be
\Psi_0[\x]=\Psi_{\text{Gauss}} \exp{\lm^2 \int\prod_{i=1}^4  dx_i~ G_4(x_1^{\D_2},x_2^{\D_2},x_3^{\D_3},x_4^{\D_3})\bchi_2(x_1)\bchi_2(x_2)\bchi_3(x_3)
\bchi_3(x_4)},
\ee
where $\Psi_{\text{Gauss}}$ comprises the Gaussian 2-point terms that give rise to the propagators above (see \eqref{eq:euc_wavefunc}) and we have shown the four-point term explicitly. We take   $G_4$ to be of the form
\be
G_4(x_1^{\D_2},x_2^{\D_2},x_3^{\D_3},x_4^{\D_3})
=  \int_{d\over2}^{{d\over2}+i\infty} {d\D\over 2\pi i} L(\D) \fourpw_{\D}^{\D_2\D_2\D_3\D_3}(x_1,x_2,x_3,x_4),
\ee
where $\lm$ is a perturbative parameter.Admittedly, this non-Gaussianity is slightly contrived since we have chosen a term that couples $\bchi_2$ to $\bchi_3$ but set other non-Gaussianities to zero. This choice is made for the purpose of obtaining a simple analytic answer. We invite the reader to generalize this to other cases.

The normal ordering in \eqref{nongaussiansetup} can be defined using the same subtractions as in the free-field case. This is sufficient to ensure the conditions of Section \ref{secfeynman} at lowest order in perturbation theory. 

\begin{figure}[h]
	\centering
	\begin{subfigure}[b]{0.35\textwidth}
	\includegraphics[width=\textwidth]{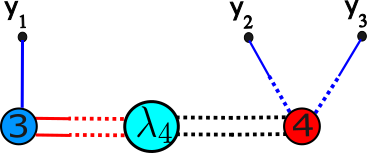}
        \caption{\label{figintfn1}}
\end{subfigure}	
\hfil
\begin{subfigure}[b]{0.35\textwidth}
\includegraphics[width=\textwidth]{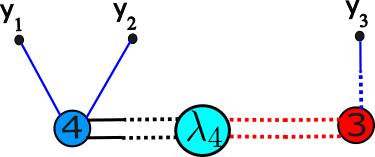}
\caption{\label{figintfn2}}
\end{subfigure}
	\caption{Feynman diagrams for the 3-point function in a 3+4 particle state built on a non-Gaussian vacuum.}
	\label{fig34lmfn1}
\end{figure}
We will compute $\llangle \psi| \x_1(y_1) \x_1(y_2) \x_1(y_3)|\psi\rrangle$. Without the non-Gaussianity shown above, this quantity would have vanished. In the presence of the non-Gaussianity, 
the two relevant diagrams are shown in Figure \ref{fig34lmfn1}. (We use blue for $\x_1$, red for $\x_2$ and black for $\x_3$).  
Denoting the value of the diagram in Figure \ref{figintfn1} by $\mathcal{F}_5$ we find
\be
\begin{split}
	\mathcal{F}_5&=\lm^2 (2!)^3 \mathfrak{c}_{\x_1\bchi_1}^2 \mathfrak{c}_{\x_2\bchi_2}^2 \mathfrak{c}_{\bchi_3\bchi_3}^2 \mathfrak{c}_{\x_1\x_1} \int \prod_{i=1}^4 dx_i 
	\g G_3^*(y_1^{S[\D_1]},x_1^{\D_2},x_2^{\D_2}) G_4(x_1^{\D_2},x_2^{\D_2},x_3^{\D_3},x_4^{\D_3}) 
	\\&\times \g G_4(y_2^{\D_1},y_3^{\D_1},x_3^{S[\D_3]},x_4^{S[\D_3]})
	\\&=\lm^2 (2!)^3 \mathfrak{c}_{\x_1\bchi_1}^2 \mathfrak{c}_{\x_2\bchi_2}^2 \mathfrak{c}_{\bchi_3\bchi_3}^2 \mathfrak{c}_{\x_1\x_1}  \int_{d\over2}^{{d\over2}+i\infty} {d\D\over 2\pi i} L(\D) \int_{d\over2}^{{d\over2}+i\infty}{d\D'\over 2\pi i} I(\D')
	\\& \times\int \prod_{i=1}^4  dx_i \g G_3^*(y_1^{S[\D_1]},x_1^{\D_2},x_2^{\D_2})
	\fourpw_{\D}^{\D_2\D_2\D_3\D_3}(x_1,x_2,x_3,x_4) \fourpw_{\D'}^{\D_1\D_1 S[\D_3]S[\D_3]}(y_2,y_3,x_3,x_4)
				\\&=\lm^2 {\mathcal{K}_5\over \abs{y_1-y_2}^{\D_1}\abs{y_1-y_3}^{\D_1}\abs{y_2-y_3}^{\D_1}}.
\end{split}
\ee
The $x_i$ integrals can be done using the usual techniques. First we write the conformal partial waves in their integral representation \eqref{eq:CPW_defn} and then use the bubble integral formula \eqref{eq:bubble integral formula}. In the final step  the numerical factor is given by
\be
\mathcal{K}_5={(2!)^3 \mathfrak{c}_{\x_1\bchi_1}^2 \mathfrak{c}_{\x_2\bchi_2}^2 \mathfrak{c}_{\bchi_3\bchi_3}^2 \mathfrak{c}_{\x_1\x_1}
	L(\dbar_1)I(\D_1) \mcalc^*_{S[\D_1]\D_2\D_2} C_{\D_2\D_2\dbar_1} C_{\D_1\D_3\D_3} C_{\D_1\D_1\D_1} C_{\dbar_1 S[\D_3]S[\D_3]} \over \left(\rotgrpvol \mu(\D_1)\right)^2}.
\ee
Note that the 3 point function has the required functional form. The multiplicative constant $\mathcal{K}_5$ depends on the interaction vertex as well as the state.

\begin{figure}[H]
	\centering
	\begin{subfigure}[b]{0.35\textwidth}
		\centering
		\includegraphics[width=\textwidth]{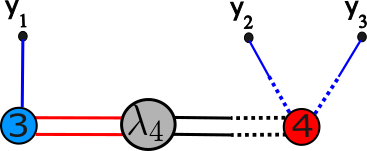}
		\caption{}
	\end{subfigure}
	\hfil
	\begin{subfigure}[b]{0.35\textwidth}
		\centering
		\includegraphics[width=\textwidth]{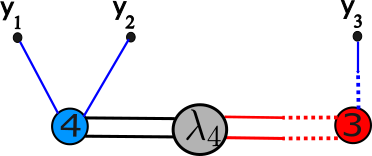}
		\caption{}
	\end{subfigure}
	\caption{Additional Feynman diagrams for the 3-point cosmological correlator in a 3+4-particle state built on an interacting vacuum \label{fig34diagconj}}
\end{figure}

Denoting the value of the diagram in Figure \ref{figintfn2} by $\mathcal{F}_6$ we have
\be
\begin{split}
	&\mathcal{F}_6=\lm^2 (2!)^3 \mathfrak{c}_{\x_1\x_1}^2 \mathfrak{c}_{\bchi_2\bchi_2}^2 \mathfrak{c}_{\x_3\bchi_3}^2 \mathfrak{c}_{\x_1\bchi_1}
	\int \prod_{i=1}^4  dx_{i} \g G_3(y_3^{\D_1},x_1^{S[\D_2]},x_2^{S[\D_2]}) 
	G_4(x_1^{\D_2},x_2^{\D_2},x_3^{\D_3},x_4^{\D_3}) 
	\\& \times \g G_4^*(y_1^{S[\D_1]},y_2^{S[\D_1]},x_4^{\D_3})
	\\&= \lm^2 (2!)^3 \mathfrak{c}_{\x_1\x_1}^2 \mathfrak{c}_{\bchi_2\bchi_2}^2 \mathfrak{c}_{\x_3\bchi_3}^2 \mathfrak{c}_{\x_1\bchi_1}
	\int_{d\over2}^{{d\over2}+i\infty} {d\D\over 2\pi i} L(\D) \int_{d\over2}^{{d\over2}+i\infty}{d\D'\over 2\pi i} I(\D')^*  \Big[
	\\& \int \prod_{i=1}^4  dx_i 
	\g G_3(y_3^{\D_1},x_1^{S[\D_2]},x_2^{S[\D_2]})  \fourpw_{\D}^{\D_2\D_2\D_3\D_3}(x_1,x_2,x_3,x_4) \left(\fourpw_{\D'}^{S[\D_1]S[\D_1]\D_3\D_3}(y_1,y_2,x_3,x_4)\right)^* \Big]
					\\&=\lm^2 {\mathcal{K}_6\over \abs{y_1-y_2}^{\D_1}\abs{y_1-y_3}^{\D_1}\abs{y_2-y_3}^{\D_1} }.
\end{split}
\ee
Here,
\be
\mathcal{K}_6={ (2!)^3 \mathfrak{c}_{\x_1\x_1}^2 \mathfrak{c}_{\bchi_2\bchi_2}^2 \mathfrak{c}_{\x_3\bchi_3}^2 \mathfrak{c}_{\x_1\bchi_1}
	L(\dbar_1) I(\dbar_1)^* \mcalc_{\D_1S[\D_2]S[\D_2]} C_{\D_2\D_2\dbar_1} \abs{C_{\D_1\D_3\D_3}}^2 C^*_{S[\D_1]S[\D_1]\dbar_1} \over 
	\left(\rotgrpvol \mu(\D_1)\right)^2}.
\ee
One can multiply by a factor of 3 to take into account other diagrams obtained after permuting $y_1,y_2$ and $y_3$. 
One also has to consider diagrams arising from the conjugate of the interaction vertex (contributed by $\Psi^*[\bchi]$). These diagrams are shown in Figure \ref{fig34diagconj}.

The calculation of these Feynman diagrams is identical to the previous calculations and we will not repeat it here. 
They evaluate to  
\[
\lm^2 {\mathcal{K}_7\over \abs{y_1-y_2}^{\D_1}\abs{y_1-y_3}^{\D_1}\abs{y_2-y_3}^{\D_1} },
\]
and 
\[
\lm^2 {\mathcal{K}_8\over \abs{y_1-y_2}^{\D_1}\abs{y_1-y_3}^{\D_1}\abs{y_2-y_3}^{\D_1} }
\]
respectively, with the constants given by
\be
\mathcal{K}_7={(2!)^3 \mathfrak{c}_{\x_1\bchi_1}^2 \mathfrak{c}_{\x_3\bchi_3}^2 \mathfrak{c}_{\x_2\x_2}^2 \mathfrak{c}_{\x_1\x_1} L(\D_1)^* I(\D_1) \mcalc^*_{S[\D_1]S[\D_2]S[\D_2]} C^*_{\D_2\D_2\D_1} \abs{C_{\dbar_1\D_3\D_3}}^2 C_{\D_1\D_1\D_1} \over \left(\rotgrpvol \mu(\D_1)\right)^2},
\ee
and
\be
\mathcal{K}_8= {(2!)^3 \mathfrak{c}_{\x_1\x_1}^2 \mathfrak{c}_{\x_2\bchi_2}^2 \mathfrak{c}_{\x_3\x_3}^2 \mathfrak{c}_{\x_1\bchi_1} L(\D_1)^* I(\dbar_1)^* |C_{\D_1\D_2\D_2}|^2 C^*_{\dbar_1\D_3\D_3} C^*_{\D_1S[\D_3]S[\D_3]} C^*_{S[\D_1]S[\D_1]\dbar_1} \over \left(\rotgrpvol \mu(\D_1)\right)^2}
\ee
So, the $O(\lm^2)$ result for the 3 point function is given by,
\be
\llangle \psi| \x_1(y_1) \x_1(y_2) \x_1(y_3)|\psi\rrangle
=3 \lm^2 {a^*b\left(\mathcal{K}_5+\mathcal{K}_7\right)+b^*a\left(\mathcal{K}_6+\mathcal{K}_8\right) \over \abs{y_1-y_2}^{\D_1}\abs{y_1-y_3}^{\D_1}\abs{y_2-y_3}^{\D_1} }.
\ee

\section{Relational observables \label{secrelational}}
In the previous sections, we have studied cosmological correlators of ``simple operators'' in simple states. We found that these correlators do not match the vacuum-expectation values of QFT operators. 
 In this section we show that in the presence of a heavy background state, which can be thought of as an observer, it is possible to construct relational observables whose correlators correspond to QFT vacuum-expectation values to good accuracy. Our construction utilizes the techniques initially developed in \cite{Papadodimas:2015xma,Papadodimas:2013jku} and recently utilized in \cite{Bahiru:2023zlc,Bahiru:2022oas,Jensen:2024dnl}. In Appendix \ref{adscftreview}, we provide a quick review of the construction of relational observables in AdS/CFT. We urge readers more familiar with that setting to scan Appendix \ref{adscftreview} to gain intuition about the construction below. The study of relational observables has a long history going back to DeWitt \cite{DeWitt:1967yk}. (See \cite{kuchar1991problem,Rovelli:2001bz,Giddings:2005id} and references in those papers for discussion from various perspectives.)  Related recent work in the AdS context may be found in \cite{Chataignier:2024eil,DeVuyst:2024pop,Fewster:2024pur,Geng:2024dbl}. The dS construction of \cite{Kaplan:2024xyk} is also related  although the objective there was to construct observables in a static patch whereas here we are interested in the late-time slice.

\subsection{Simple observables}
We will construct operators whose expectation values and variance will match those of simple QFT observables in the vacuum. By simple QFT observables, we mean  low-point polynomials in the light fields of the theory that are localized to a given patch of the late-time slice. 

\begin{figure}[!ht]
\begin{center}
\includegraphics[width=0.3\textwidth]{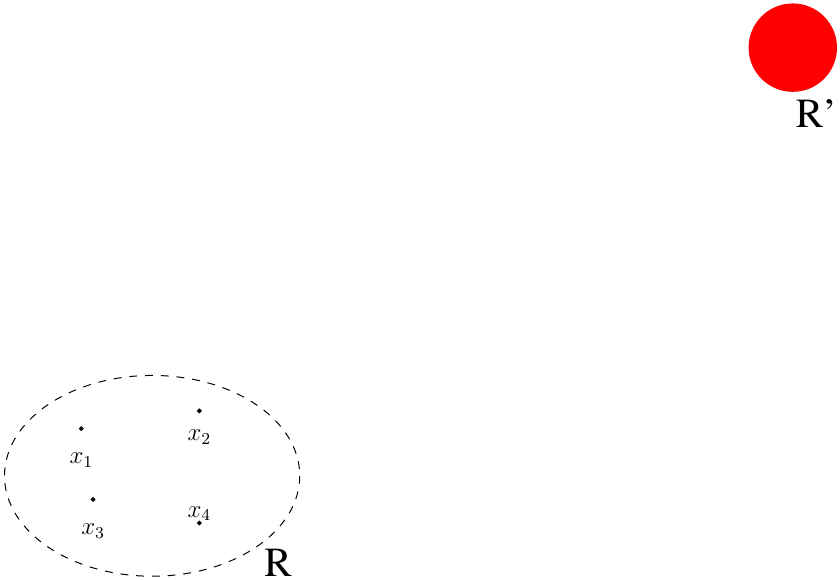}
\caption{A heavy background excitation (shown as a red blob) in the region $R'$ allows us to define relational observables that mimic QFT observables in the region R far from the background excitation. \label{relationalfig}}
\end{center}
\end{figure}
More explicitly, let $\chi(x)$ be the late-time value of a matter field. As above, this field can be in the principal series or the complementary series. Consider the set of QFT observables
\be
\alset = \text{span}\{\chi(x_1), \chi(x_1) \chi(x_2), \ldots \chi(x_1) \ldots \chi(x_n) \},
\ee
where we impose the cutoff $n < n_c$. The cutoff cannot taken to be ``too large'' and we make this more precise below.  Furthermore we take all the points $x_i$ to be localized in a given region $R$ on the late-time slice. See Figure \ref{relationalfig}.

To the extent that we can ignore this cutoff, the set $\alset$ is also an algebra in that one can multiply elements of $\alset$ to obtain other elements in $\alset$. This is precisely the sense in which local bulk algebras appear in the boundary theory --- a discussion that was initiated in \cite{Banerjee:2016mhh} and has later been extended in the language of von Neumann algebras following \cite{Leutheusser:2022bgi}. In this paper, we do not explore this algebraic structure further.

We denote the QFT expectation values of these operators in the Euclidean vacuum state by
\be
\langle a \rangle \equiv \langle 0 | a | 0 \rangle_{\qftn},
\ee
with $a \in \alset$. 

\paragraph{Statement of objective.}
The setup above allows us to precisely state the objective of our construction. We seek to construct a de Sitter invariant state, $\psiobs$, and a map from $a \rightarrow \hat{a}$ that maps each element of $a \in \alset$ to a de Sitter invariant operator $\hat{a}$ with the property that
\be
\left(\psiobs, \hat{a} \psiobs \right) = \langle a \rangle.
\ee
The relation above is expected to hold at leading order in perturbation theory in $G_N$. The map will be state dependent in that it will depend on the background state $\psiobs$.

\subsection{QFT background state}
Next, we require a quantum-field theory state that can serve as an appropriate background. Upon group averaging this state will give rise to a dS invariant ``observer''. 

The QFT background state that we require must have two important properties. First, we require the state to have the property that
\be
\label{statelikevac}
\langle \psiback | a | \psiback \rangle \approx \langle 0| a | 0 \rangle,
\ee
where $|0 \rangle$ is the Hartle-Hawking state. Here the $\approx$ sign means that the correlators are equal to high accuracy. This accuracy will be one of determinants of the accuracy of our construction. 

Physically this condition is easy to achieve. Recall that the elements of $\alset$ are all centered in a particular region $R$. If the background state has excitations in another part of the late-time slice then we expect that the presence of the excitations will not affect probes in $R$. 

Next recall that, on the quantum field theory Hilbert space, the de Sitter group can be realized via unitary operators, $U$, that map a field operator on the late-time slice to another operator also on the late-time slice. We require the QFT background state to satisfy the property that if $a \in \alset$ then if $U$ is outside of a small set, $\Sigma \subset \confgrp$ near the identity we have
\be
\label{rapidfalloff}
|\langle \psiback | U a | \psiback \rangle| \ll 1, \qquad U \notin \Sigma.
\ee

The basic idea is to consider a state made up of a large number of well-localized elementary excitations. The action of any element of the de Sitter group (except for elements that are parametrically close to the identity) moves these elementary excitations around. This produces a state that is orthogonal to the original state.

The state must be ``heavy'' since it must have a large number of excitations to satisfy \eqref{rapidfalloff}. This does not require the state must have enough localized energy density to backreact substantially. As explained in  \cite{Jensen:2024dnl} even a single virion is heavy enough to meet the condition \eqref{rapidfalloff} but too small to warp the spacetime appreciably. 

 In Appendix \ref{subsecobs}, we provide an explicit example of a state satisfying  \eqref{rapidfalloff} and \eqref{statelikevac} in the free-field approximation. We do not see any obstacle to finding a state that satisfies these properties in the interacting theory.  Going forward, we will simply assume that such a state has been found.

\subsection{Auxiliary smeared quantities \label{secauxsmeared}}
The last technical step that is required is to define some auxiliary smeared versions of the little Hilbert space and operators.  Their utility will be apparent below.

First, we choose a set $\Sigma' \subset \confgrp$ which is also centered about the identity. We want $\Sigma'$ to have the property that it is much larger than $\Sigma$ but still correspond to small elements of the conformal group in the sense that
\be
\label{invarianceu}
\langle a U b U^{\dagger} \rangle \approx \langle a b \rangle, \qquad U \in \Sigma'.
\ee
In what follows we will neglect terms that are of size ${\text{vol}(\Sigma) \over \text{vol}(\Sigma')}$.

We now define a smeared little Hilbert space about the original QFT state as
\be
{\cal H}_{\text{sm}}  = \text{span} \{ a U |\psiback \rangle, \qquad a \in \alset, U \in \Sigma'\}.
\ee
Note that, by the Reeh-Schlieder property of ordinary quantum fields in de Sitter space \cite{Bros:1995js}, this space is dense in the space
\be
{\cal H}'_{\text{sm}} = \text{span} \{U a |\psiback \rangle, \qquad a \in \alset, U \in \Sigma' \}.
\ee
We now define the projector $P$ to be the projector onto the space ${\cal H}_{\text{sm}}$.

Next, for every operator $a$ we define a ``smeared'' operator
\be
\bar{a} = {1 \over \text{vol}(\Sigma')} \int_{\Sigma'} U a U^{\dagger} d U,
\ee
If $a$ contains some local operators then the conformal transformations move these operators about by a small amount. To the extent that all elements of $\Sigma'$ are close to the identity, we expect that insertions of $\bar{a}$ are
indistinguishable from insertions of $a$ within expectation values and we will assume this below.

Finally, we define a slightly smeared version of the state $|\psiback \rangle$ as
\be
|\psibacksm \rangle = {{\cal N} \over \sqrt{\text{vol}(\Sigma')}} \int_{\Sigma'} d U U |\psiback \rangle,
\ee
where the normalization constant ${\cal N}$ is chosen as
\be
\label{normdef}
{\cal N}^{-2} = \int \langle \psiback | U |\psiback \rangle d U.
\ee
This ensures that  
\be
\langle \psibacksm | \psibacksm \rangle = {{\cal N}^2 \over \text{vol}(\Sigma')} \int_{\Sigma'} d U \int_{\Sigma'} d V \langle  \psiback | U V  |\psiback \rangle  = 1.
\ee
In the last line we note that for any $V \in \Sigma'$ we must have $U V \in \Sigma$ for the inner product to be appreciable. Therefore,  the inner product vanishes unless $U$ is in a small open set about $V^{\dagger}$. Hence, up to edge-effects of size ${\text{vol}(\Sigma) \over \text{vol}(\Sigma')}$ we can extend the integral over $U$ to the entire group. We then use the invariance of the Haar measure to redefine $U \rightarrow U V$ so that the integrand is independent of $V$. The integral over $V$ cancels the factor of $\text{vol}(\Sigma')$ in the denominator and the remaining integral over $U$ cancels the factor of ${\cal N}^2$.

Physically, we can think of the state $|\psibacksm \rangle$ as a superposition of states where the background density has been ``moved around'' a little by conformal transformations. Using the same steps as above, we see that
\be
\label{smearedassmeared}
\langle \psibacksm | a | \psibacksm \rangle = {\cal N}^2 \int \langle \psiback | U a |\psiback \rangle d U.
\ee
Since we are still focused on insertions far away from the background, it is natural to expect that
\be
\label{smearedalsovac}
\langle \psibacksm | a |\psibacksm \rangle = \langle a \rangle.
\ee
This leads us to the conclusion that
\be
\label{normcorrectwitha}
{\cal N}^2 \int \langle \psiback | U a |\psiback \rangle d U = \langle a \rangle
\ee
This shows that the normalization factor chosen to normalize the state in \eqref{normdef} is also the correct normalization factor to use in \eqref{normcorrectwitha}. The utility of defining $|\psibacksm \rangle$ is that it allows us to give a physical proof of this result. If one has a more macroscopic description for $|\psiback \rangle$, then it might be possible to prove \eqref{rapidfalloff} and \eqref{normcorrectwitha} directly as, for instance, was done in section 7.6 of \cite{Papadodimas:2015jra}. We will not need $|\psibacksm \rangle$ again.

\subsection{Observer state and relational observables}
We now have all the ingredients to  construct ``relational observables'' whose correlators in an appropriate ``observer state'' are like correlators of simple QFT operators in the Hartle-Hawking state.

For any QFT operator $a \in \alset$, we define 
\be
\label{hata}
\hat{a} = {1 \over \text{vol}(\Sigma')} \int V   a P V^{\dagger} dV.
\ee
We also define the state
\be
|\psiobs \rangle = {\cal N}_1  \int d U U |\psiback \rangle.
\ee
with ${\cal N}_1 = {\cal N} \left(\text{vol}(\sodminone) \right)^{-{1 \over 2}}$. It is manifest that $\hat{a}$ is a dS invariant operator and $|\psiobs \rangle$ is a dS-invariant state since both the operator and the state are group averaged. 

We also note that the state $|\psiobs \rangle$ is normalized correctly since
\be
\begin{split}
\left( \psiobs, \psiobs \right) &= {{\cal N}^2} {1 \over \text{vol}(\confgrp)} \int \langle \psiback | U^{\dagger} V |  \psiback \rangle d U d V \\ 
&= {\cal N}^2 \int \langle \psiback | V |\psiback \rangle d V = 1,
\end{split}
\ee 
In going from the first line to the second line, we use the invariance of the Haar measure to redefine $V \rightarrow U^{\dagger} V$. the integral over $U$ then cancels the volume of $\confgrp$ leaving behind an integral that manifestly cancels the normalization factor.

We now show that expectation values of the operators $\hat{a}$ on the state $|\psiobs \rangle$ yield the vacuum expectation values $\langle a \rangle$. First note that 
\be
\label{ahatonstate}
\begin{split}
&\hat{a} |\psiobs \rangle = {{\cal N}_1 \over  \text{vol}(\Sigma')} \int V   a P V^{\dagger}  U |\psiback \rangle \,  dV  d U \\
&= {{\cal N}_1 \over \text{vol}(\Sigma')} \int   U W   a P W^{\dagger} |\psiback \rangle \, dW  d U \\
&={\cal N}_1  \int  U \bar{a} |\psiback \rangle \,  d U
\end{split}
\ee
Here we first defined $V = U W$ and so $V^{\dagger} = W^{\dagger} U^{\dagger}$. There is no Jacobian factor due to the invariance of the Haar measure. Next we note that the projector restricts the range of the $W$ integral to $W \in \Sigma'$. This restriction is {\em not} exact but, by \eqref{rapidfalloff},  it is valid up to terms of size $\Or[{\vol{\Sigma} \over \vol{\Sigma'}}]$. In words,  acting with the hatted operator on the group-averaged state is the same as acting with the smeared operator on the original state and then group averaging the result.

Therefore, 
\be
\label{relationalproof}
\begin{split}
\left( \psiobs , \hat{a} \psiobs \right) &= {{\cal N}^2 \over \text{vol}(\confgrp)}  \int d U d U' \langle \psiback | (U')^{\dagger} U \bar{a} | \psiback \rangle  \\
&= {\cal N}^2   \int d U \langle \psiback |  U \bar{a} | \psiback \rangle \\
&=  \langle a \rangle.
\end{split}
\ee
In the first step here, we redefined $U \rightarrow (U')^{\dagger} U $. There is no Jacobian due to the invariance of the Haar measure, and the integral over $U'$ simply cancels the volume of the conformal group.  We then used the relation \eqref{smearedassmeared} and approximated the expectation value of the smeared operator by the original operator.

Repeating the manipulations above shows that the map $a \rightarrow \hat{a}$ preserves the product structure of $\alset$. Namely,
\be
\hat{a}_1 \hat{a}_2 \ldots \hat{a}_n |\psiobs \rangle = {\cal N}_1  \int  U \bar{a}_1 \bar{a}_2 \ldots \bar{a}_n |\psiback \rangle \,  d U,
\ee
and therefore,
\be
\left( \psiobs , \hat{a}_1 \ldots \hat{a}_n \psiobs \right) = \langle a_1 \ldots a_n \rangle,
\ee
As a special case, this shows that the variance of $\hat{a}$ in the state $|\psiobs \rangle$ coincides with the variance of $a$ in the vacuum. This provides evidence that  the quantum probabilities of various outcomes obtained by measuring $\hat{a}$ in the state $|\psiobs \rangle$ are close to those of outcomes obtained by measuring $a$ in $|0 \rangle$.

This completes the construction of relational observables whose expectation values in a particular state match the vacuum expectation values of simple operators.

The proof above also shows why the smeared quantities in Section \ref{secauxsmeared} were necessary. The key relation \eqref{ahatonstate} is valid up to terms of order ${\text{vol}(\Sigma) \over \text{vol}(\Sigma')}$. If one tries to make the projectors too sharp, by reducing $\text{vol}(\Sigma')$ then this invalidates this relation. Some more comments may be found in Appendix \ref{apprelationaldetails}.

\subsection{Comments on the construction}
We would like to make some comments on the construction presented in the previous subsection.
\begin{enumerate}
\item
A salient feature of the observables is that they are tailored to a specific background state. This state-dependence is manifest in the projectors that appear in \eqref{hata}. In particular, the same observables would not make sense in another state. This can be justified as follows. A physical observer finds themselves in a particular state of the Universe. In that state, they choose to describe the physics around them using observables whose correlators behave like simple local operators. On the other hand, state dependence raises broader questions and potential paradoxes that have been explored in the context of AdS/CFT \cite{Papadodimas:2015jra,Marolf:2015dia,Raju:2016vsu}. (See \cite{Raju:2020smc} for a review.) So this feature of the construction deserves further attention.
\item
This construction is valid to leading order in $G_N$ perturbation theory. The validity of the construction relies on \eqref{rapidfalloff} but the size of $\Sigma$ is typically perturbative in $G_N$. (See Appendix \ref{apprelationaldetails} for an example.) It is an open technical problem to extend this construction to higher-orders in perturbation theory. We do not see any obstacle to this extension although we would like to caution the reader against premature claims in the literature which suggest that such an extension is trivial. Such an extension is important to understand gravitational corrections to these relational observables. Some of the technical issues are explained in \cite{Jensen:2024dnl}.
\item
While it might be possible to improve  \eqref{relationalproof} so that it is valid at higher orders in perturbation theory, it can never be made exact. The accuracy of the approximation is controlled by features of the background state. This tells us that if one makes precise-enough measurements to detect the background, then the nonlocal features of these cosmological observables eventually become evident. Therefore, the principle of holography of information is valid as an exact statement, even in the presence of a heavy background.
\end{enumerate}

\section{Discussion}
In this paper, we studied cosmological correlators in the Hilbert space of gravitationally constrained de Sitter states in the $G_N \to 0$ limit.  Even in this limit,  it is necessary to impose the global part of the constraints and this forces all states to be de Sitter invariant. The Hilbert space is obtained by group averaging ordinary QFT states and defining the inner product to be the QFT inner product divided by the volume of the de Sitter group.

We utilized a convenient representation of such states that was obtained in \cite{Chakraborty:2023yed} by taking  the nongravitational limit of solutions to the WDW equation in asymptotically de Sitter space: de Sitter invariant states are obtained by smearing ordinary field excitations above the Euclidean vacuum with an appropriate conformally covariant function. 

Gauge-invariant operators can also be obtained by group averaging QFT operators.  Group averaging a product of ordinary quantum-field operators leads to a gauge-invariant operator that is represented by smearing the product of operators with a conformally covariant function.  The gauge-fixed version of such an observable was termed a ``cosmological correlator'' in \cite{Chakraborty:2023los}. 

In Section \ref{secfeynman}, we formulated Feynman rules to compute cosmological correlators in de Sitter invariant states. We also explained how to remove a simple set of group-volume divergences that might appear in some observables. This leads to some necessary conditions for states to have finite norm, and for observables to have finite expectation values.

In Section \ref{secsample}, we evaluated cosmological correlators in some simple examples. A striking feature of our result is that cosmological correlators take on nontrivial values even when the Euclidean vacuum is represented by a Gaussian wavefunction. We utilized techniques from harmonic analysis on the conformal group to obtain explicit expressions for some cosmological correlators. The final expressions are all conformally covariant, which is expected from the analysis of the symmetries of conformal correlators performed in \cite{Chakraborty:2023los}. 

The fact that cosmological correlators differ from QFT correlators in the vacuum, and can be nontrivial even when the vacuum is trivial is to be expected. Any normalizable state must have excitations on top of the vacuum and these excitations also contribute to the cosmological correlator. However, in the gravitational case, this contribution is conformally invariant. In a nongravitational QFT, cosmological correlators in normalizable excited states are never conformally invariant unless the excitation doesn't contribute to the correlator at all. 

Nevertheless, one might ask whether it is possible to find any observables whose expectation values approximate quantum-field theory cosmological correlators in the Euclidean vacuum. We answer this question affirmatively in Section \ref{secrelational}. The trick is to consider a heavy background state obtained by group averaging a QFT-state that is localized in one part of the late-time slice. We then define relational observables with respect to this background state. This construction utilizes techniques previously developed to study quasilocal observables in AdS/CFT in \cite{Papadodimas:2015xma,Papadodimas:2015jra,Bahiru:2022oas,Bahiru:2023zlc,Jensen:2024dnl}.

The construction of Section \ref{secrelational} does not violate the principle of holography of information. This is because while the observables constructed in \ref{secrelational} approximate expectation values in the Euclidean vacuum to leading order in perturbation theory, a sufficiently precise examination of a sufficiently high-point cosmological correlator would cause the construction to break down and would reveal the presence of the heavy background. 

The holographic properties of gravity are somewhat analogous to unitarity in ordinary quantum field theories. Unitarity is an exact statement at the nonperturbative level and is important for the consistency of the theory. Unitarity can also be verified for light observables about empty space, such as the S-matrix of a few incoming and a few outgoing particles. Nevertheless, in the presence of a heavy background when one coarse-grains the set of observables, physics appears dissipative. This is a feature, not a bug, because it helps to reconcile the principle of unitarity with our mundane experience of dissipation.

The same regimes apply to the principle of holography of information. In principle, gauge-fixed correlators in any open set on the late-time slice provide complete information about the state. The calculations of \ref{secsample} show that this result has perturbative signatures; in perturbative states with a small number of excitations, low-point cosmological correlators are sensitive to the excitations. Nevertheless, in the presence of a heavy background, when one coarse-grains the set of observables, several distinct states can resemble the Hartle-Hawking state. This is also a feature, and not a bug, because it helps to reconcile holography with our mundane experience of local physics.

The relational observables of Section \ref{secrelational} are related to the observables that can be defined in the presence of an observer \cite{Chandrasekaran:2022cip} or in the presence of a timelike boundary \cite{Silverstein:2024xnr}. A benefit of the relational perspective is that the observer arises naturally from the property of an excitation in a closed universe and one does not need to puncture the spacetime; the cost that one must pay is that the construction is technically more elaborate.

This study raises several interesting issues that we hope to address in future work. An immediate extension is to improve the analysis of Section \ref{secsample} to incorporate loops.  This will require us to renormalize UV and IR divergences \cite{Senatore:2009cf,Chowdhury:2023arc,Bertan:2018khc,Bzowski:2023nef} and incorporate the effects of anomalous dimensions in the fields. From a more conceptual perspective, certain singularities of  QFT correlators are fixed by a ``flat space limit'' \cite{Maldacena:2011nz,Raju:2012zr,Arkani-Hamed:2015bza}. This constrains the correlator everywhere \cite{Baumann:2021fxj}. Is there a similar structural constraint that the cosmological correlators in the gravitational Hilbert space must obey in the perturbative limit?

Another relevant technical problem is to extend the construction of Section \ref{secrelational} to higher orders in perturbation theory. The construction so far yields relational observables that approximate QFT correlators that are smeared by the action of elements of the conformal group drawn from a small open set near the identity. But this smearing is not nonperturbatively small. This means that the relational observables differ from QFT correlators at higher-orders in perturbation theory. Can one systematically evaluate these effects? Does the short-distance structure of relational observables obey the Hadamard condition at higher-orders in perturbation theory and yield a finite local stress-tensor using the usual methods of QFT in curved spacetime  \cite{birrell1984quantum}?

Next, there has been considerable recent work on the formal properties of quantum field theory in de Sitter space  \cite{Penedones:2023uqc,Anninos:2023lin,Anous:2020nxu,Loparco:2023rug,Hogervorst:2021uvp,Polyakov:2012uc,Gorbenko:2019rza}. Are these analyses modified if, instead of bare quantum-field observables, one uses the dressed observables presented in Section \ref{secrelational}? 

In this paper, our calculations were performed within an infinite-dimensional Hilbert space. Nonperturbative corrections to the inner product may cause the true Hilbert space to be finite dimensional \cite{fischler:talk,Banks:2000fe,Witten:2001kn,Banks:2001yp,Arenas-Henriquez:2022pyh,Arias:2019pzy,Banks:2018ypk,Banks:2020zcr,Coleman:2021nor}. We do not expect the perturbative results presented here to be affected by such corrections.

There has also been recent discussion on the possibility that the Hilbert space of a closed Universe might be one-dimensional  \cite{McNamara:2020uza,Usatyuk:2024isz,Harlow:2025pvj}. Since our analysis is perturbative, it does not shed much light on this issue.  However, some of the arguments given for a one-dimensional Hilbert space are not very persuasive. For instance, it might naively appear that the Hartle-Hawking state is the only normalizable state that satisfies the constraints. But, as we have discussed, an alternative quantization gives rise to a rich Hilbert space. Moreover, the observation that the entanglement wedge of a point is the entire spacetime is consistent with the principle of holography of information: gauge-fixed observables in an infinitesimal open set of the late-time slice are sufficient to tell us about the entire spacetime when it is in a pure state. Arguments based on the gravitational path integral are somewhat inconclusive because the rules are ill defined. 

On the other hand, the construction of \cite{Chakraborty:2023yed} assumes the existence of  continuous families of solutions to the Wheeler-DeWitt equation; the excited states that appear in this paper are obtained by differentiating those solutions with respect to the continuous parameters. This assumption is valid in effective field theory but perhaps subtle UV effects disallow such families. Moreover, it is true that a closed Universe always remains in one state since, in the absence of external sources, it is impossible to transition to another state. So, perhaps an observer in one state can never learn about the presence of other states in the Hilbert space!

These considerations  lead to deeper questions about the nature of observations in a cosmological setting. The standard von Neumann theory of measurements envisions a modification of the Hamiltonian of the observed system that couples it to a pointer. Clearly, this does not make sense in the context of gravity, where a local modification of the Hamiltonian is inconsistent with the constraints. So a complete description of measurements is only possible in an autonomous setup where the observer is part of the system. Presumably, this more-sophisticated theory will also elucidate how the state dependent observables of Section \ref{secrelational} are automatically selected by such an observer in the semiclassical limit.

\section*{Acknowledgments}
We are grateful to Tarek Anous, Joydeep Chakravarty, Chandramouli Chowdhury, Paolo Creminelli, Victor Godet,  Diksha Jain, Sadra Jazayeri, Alok Laddha, R. Loganayagam, Mehrdad Mirbabayi, Enrico Pajer, Priyadarshi Paul, Ashoke Sen and Sandip Trivedi for helpful discussions. S.R. and A.H are grateful to ICTP (Trieste) for hospitality and support from the ICTP Simons associates program while this work was being completed.  The work of T.C. is supported by the Infosys Endowment for the study of the Quantum Structure of Spacetime. Research at ICTS-TIFR and TIFR-Mumbai is supported by the Department of Atomic Energy, Government of India, under Project Identification Nos. RTI4001 and RTI4002 respectively. 

\appendix
\section*{Appendix}
\section{Cosmological correlators as gauge-invariant expectation values \label{seccosmgaugeinv}}
In this Appendix, we explain how to represent cosmological correlators as expectation values of gauge-invariant operators. It is sufficient to study cosmological correlators corresponding to simple monomials of the field operators since more general correlators can be written as linear combinations of these.  In the nongravitational limit, we use the simple relation
\be
\langle \Psi | \chi(z_1) \ldots \chi(z_n) | \Psi \rangle_{\qftn} = \left(\Psi, O \Psi \right),
\ee
where
\be
\label{gaugeinvop}
O = {1 \over \text{vol}(\sodminone)} \int U \chi(z_1) \ldots \chi(z_n) U^{\dagger} d U,
\ee
where $d U$ is the Haar measure on the set of unitary operators, $U$, that implements conformal transformations in the QFT Hilbert space.

We recall that if $g$ is an element of the conformal group, these operators act on fields via
\be
\label{utransfields}
U \chi(\widetilde{x}) U^{\dagger} = \Lambda(\widetilde{x}, g)^{\Delta} \chi(x),
\ee
where
\be
x = g \widetilde{x}; \qquad \Lambda(\widetilde{x}, g) = \left|{\partial x \over \partial \widetilde{x}} \right|^{1 \over d}.
\ee
Here, we use $g \widetilde{x}$ to denote the action of the conformal transformation $g$ on $\widetilde{x}$. We caution the reader that the notation is slightly different from \cite{Chakraborty:2023yed}.

Therefore, we can write
\be
O = {1 \over \text{vol}(\sodminone)} \int d g \prod_{i} \Lambda(g, z_i)^{\Delta} \chi(g z_i),
\ee
where $d g$ is the Haar measure on the conformal group. 
Inserting an identity in the form
\be
1 = \int \prod_i \delta(x_i - g z_i) d x_i.
\ee
We see that we can write
\be
\label{gaugeinvopsm}
O =  \int {\cal G}(\vec{x}, \vec{z}) \prod_i \chi(x_i) d x_i,
\ee
where
\be
\label{smearingformula}
{\cal G}(\vec{x}, \vec{z}) = {1 \over \text{vol}(\sodminone)} \int \prod \delta(x_i - g z_i) \Lambda(g, z_i)^{\Delta} d g.
\ee
While this provides an explicit expression for ${\cal G}$ in terms of an integral over the conformal group, it is difficult to perform this integral by brute force. The objective of the calculations below is to identify the smearing function ${\cal G}$ in various case. We proceed in three steps. First, we identify the conformal transformation properties of ${\cal G}$ under transformations of $x_i$ and $z_i$. Next, we use this to constrain the form that ${\cal G}$ can take. Finally, we provide explicit expressions for ${\cal G}$ in the case of 3-point and 4-point cosmological correlators and explain how these calculations can be generalized to higher-point correlators.

\paragraph{\bf Transformation of the smearing function under conformal transformations.}
We first show that under a conformal transformation of the points $x_i$
\be
\label{transgx}
{\cal G}(\gamma \vec{x}, \vec{z}) = \Lambda(\gamma, x_i)^{-\dbar} {\cal G}(\vec{x}, \vec{z}).
\ee
This is also necessary to ensure that the operator as written in the form \eqref{gaugeinvopsm} is invariant under conformal transformations so that the transformation of ${\cal G}$ compensates for the transformation of the fields.

We see that
\be
\delta(\gamma x_i - g z) = \delta(x_i - \gamma^{-1} g z_i) \Lambda(\gamma, x_i)^{-d},
\ee
and, in the presence of the delta functions,
\be
\Lambda(g, z_i) = \Lambda(\gamma, x_i)  \Lambda(\gamma^{-1} g, z_i).
\ee
The required property \eqref{transgx} then follows by substituting the relations above into \eqref{smearingformula} and using the invariance of the Haar measure under left multiplication.

Using a similar argument, one may also check that under a conformal transformation of the points $z_i$, we have
\be
{\cal G}(\vec{x}, \gamma \vec{z}) = \Lambda(\gamma, x_i)^{-\Delta} {\cal G}(\vec{x}, \vec{z}).
\ee

\paragraph{\bf General form of the smearing function.}
The delta functions in \eqref{smearingformula} imply that the cross ratios of the $x_i$ must necessary be the same as the cross ratios of the $z_i$. This together with the transformation of ${\cal G}$ under independent conformal transformations of $x_i$ and $z_i$ derived above leads to the conclusion that one can write
\be
\label{formsmearing}
{\cal G}(\vec{x}, \vec{z}) = C_1(x_i) C_2(z_i) {\cal C}(\vec{u}^{z}) {\cal D}(\vec{u}^{x}, \vec{u}^{z}).
\ee
Here $C_1(x_i)$ transforms as  a conformal correlator of operators of dimension $\dbar$, $C_2(x_i)$ transforms as a conformal correlator of operators of dimension $\Delta$ and ${\cal C}$ is an arbitrary function of the cross rations made of the $z_i$. We denote the list of cross ratios by $\vec{u}^{z}$. Finally ${\cal D}$ is a product of delta functions that sets $\vec{u}^{z}$ to the cross ratios made out of the $x_i$, which we denote by $\vec{u}^{x}$. ${\cal D}$ only appears when $n > 3$.

Provided that the number of points, $3 < n \leq d + 2$, we can simply set
\be
{\cal D} = \prod_{\alpha} \delta(u^z_\alpha - u^x_\alpha),
\ee
where the product runs over all cross ratios indexed by $\alpha$.  However, when $n > d+2$ some of the cross ratios are redundant, in which case only a smaller number of delta functions can be imposed explicitly.

The form \eqref{formsmearing} has some redundancy since ${\cal C}$ can be absorbed into $C_2$. We have written it as above so that $C_1$ and $C_2$ can be chosen as per convenience and the task then reduces to finding ${\cal C}$.

Before we explain how to fix ${\cal G}$ explicitly,  we note that the relation \eqref{gaugeinvop} defines a many-to-one map between gauge-fixed operators and gauge-invariant operators i.e. multiple distinct monomials on the right hand side of \eqref{gaugeinvop} might get mapped to the same gauge invariant operator. For instance, if one acts on a monomial with any unitary from the conformal group that transforms the fields according to \eqref{utransfields} one obtains a distinct monomial. However, the image of this new monomial under the map \eqref{gaugeinvop} is the same as that of the previous monomial. 

Next, the following formula, derived in \cite{Chakraborty:2023yed}, is helpful. Let us use the notation $\vec{x'}$ to denote the list of $x$ positions with the first three fixed $x_1 = z_1, x_2 = z_2, x_3 = z_3$: $\vec{x}' \equiv (x_1 = z_1, x_2 = z_2, x_3 = z_3, x_4 \ldots x_n)$.  Then
\be
\label{ansatzg}
{1 \over f} \int d U d\vec{x}'U {\cal G}(\vec{x}', \vec{z}) \prod_{i} \chi(x_i') U^{\dagger} = \int d \vec{x} {\cal G}(\vec{x}, \vec{z}) \prod_i \chi(x_i),
\ee
with
\be
\label{fdef}
f = \text{vol(SO(d-1))} (|x_{12}'| |x_{23}'||x_{13}'|)^{-d}.
\ee

The task at hand is to deduce a smearing function ${\cal G}$ consistent with \eqref{formsmearing} so that the left hand side reduces either to the desired monomial {\em or} one of its images under the action of a unitary. This is most conveniently done by choosing an appropriate ansatz for ${\cal G}$ and fixing parameters in the ansatz. We now do this explicitly for the case of 3-point and 4-point cosmological correlators and indicate the generalization to higher points.

\subsection{3-point cosmological correlator}
For the case of the 3-point cosmological correlator we see that \eqref{formsmearing} forces a specific form on ${\cal G}$
\be
{\cal G}(\vec{x}, \vec{z}) = {1 \over |x_{12}|^{\dbar} |x_{1 3}|^{\dbar} |x_{2 3}|^{\dbar} |z_{12}|^{\Delta} |z_{1 3}|^{\Delta} |z_{2 3}|^{\Delta}},
\ee
and the overall constant is fixed using \eqref{ansatzg}. This case is simple because there are no cross ratios and so ${\cal C}$ and ${\cal D}$ cannot appear at all. 

\subsection{4-point cosmological correlator}
For the case of the 4-point correlator we make an ansatz of the form
\be
\label{ansatz4pt}
{\cal G}(\vec{x}, \vec{z}) = {1 \over \prod_{i < j} |x_{i j}|^{2 \dbar \over 3} |z_{i j}|^{2 \Delta \over 3}} \delta(u_x - u_z) \delta(v_x - v_z) h(u_z, v_z),
\ee
where $h$ is a smooth function of 
\be
u_z = {|z_{4 1}| |z_{2 3}| \over |z_{4 3}| |z_{1 2}|}; \qquad v_z = {|z_{4 2} |z_{1 3}| \over |z_{4 3}| |z_{1 2}|}.
\ee

$h$ can be fixed using \eqref{ansatzg} by demanding that the smearing function  reduce the restricted integral over $x_4$ to the product $\chi(z_1) \ldots \chi(z_4)$ or one of its conformal images. Moreover, since $h$ depends only on the cross ratios we can fix it in the special case where $z_1 = 0, z_2 = 1, z_3 = \infty$.  In this setting $u_z = |z_4|$ and $v_z = |1 - z_4|$. 

For this choice with $x_1, x_2, x_3$ set to $z_1, z_2, z_3$ the prefactors above (including the contribution of $f$)  become
\be
(|z_{12}| |z_{23}||z_{13}|)^{d} {1 \over \prod_{i < j} |x_{i j}|^{2 \dbar \over 3} |z_{i j}|^{2 \Delta \over 3}} = {1 \over u_z^{2 d \over 3}} {1 \over v_z^{2 d \over 3}}.
\ee
We find that  $h$ must be
\be
\begin{split}
h &= { u_z^{2 d \over 3} v_z^{2 d \over 3} \over  \int d x_4 \delta(|x_4| - u_z) \delta(|1-x_4| - v_z)} \\
&= {2^{d-3} u_z^{2 d \over 3} v_z^{2 d\over 3} \over  u_z v_z \Omega_{d-2} \left[\left(u_z + v_z + 1 \right) \left(u_z + v_z - 1 \right) \left(-u_z + v_z + 1 \right) \left(u_z - v_z + 1 \right)\right]^{d-3 \over 2}}.
\end{split}
\ee

Substituting this value of $h$ into \eqref{ansatz4pt} yields the correct smearing function for the four-point cosmological correlator. This answer holds for arbitrary values of $z_i$ and not just for the special values that we chose above to fix $h$. 

\subsection{Higher-point cosmological correlators}
In principle the procedure outlined above works for an arbitrary point correlation function. However, when the number of points is high enough some of the cross ratios become redundant. In this case, the delta functions that appear in the smearing function must be restricted to the independent cross ratios. 

We naively obtain ${n (n-3) \over 2}$ cross ratios from $n$ points. But for large $n$, this is an overestimate and not all cross ratios are independent. Each point has $d$ coordinates and therefore $n$ points have $n d$ degrees of freedom. Of these conformal transformations can at most remove $\text{dim}(SO(d,1)) = {(d+1)(d+2) \over 2}$ degrees of freedom. Therefore, for large $n$ the number of independent cross ratios is given by $n d - \text{dim}(SO(d,1))$. The two formulas cross over at $n = d+2$ and, in general, the number of independent cross ratios is given by \cite{ferrara1972covariant}
\be
N_{\text{ind}} = \begin{cases} {n (n - 3) \over 2} & n \leq d+2 \\ 
n d - {(d+1)(d+2) \over 2} & n > d+2 \end{cases}.
\ee

In general, one must choose an ansatz with delta functions that set $N_{\text{ind}}$ cross ratios of $x$ equal to those of $z$. The precise smearing function can then be obtained, as above, after computing the volume of the compact space left undetermined by these delta functions. 

\section{More details on relational observables \label{apprelationaldetails}}
\subsection{Review of relational observables in AdS/CFT \label{adscftreview}}
In this Appendix, we provide a lightning review of the construction of relational observables in AdS/CFT. The construction we present below was first presented in \cite{Papadodimas:2015xma,Papadodimas:2015jra} to obtain operators in the interior of the eternal black-hole that had a specified commutator with the boundary Hamiltonian. It is closely related to the construction of ``mirror operators'' in the single-sided black hole \cite{Papadodimas:2013wnh,Papadodimas:2013jku}.   The same construction was recently applied to non-black hole states in  \cite{Bahiru:2022oas,Bahiru:2023zlc} to obtain operators that commute with the boundary Hamiltonian.  The construction was slightly refined in  \cite{Jensen:2024dnl} where it was shown that the relational observables obtained via this method take on the form of a crossed product that has been studied recently in \cite{Witten:2021unn,Chandrasekaran:2022cip}. The appeal of this approach is that one does not need to introduce an ``observer''  and the properties of the background states naturally give rise to an autonomous notion of time. Second, this approach shows that relational observables confined to a compact region are only approximate; their existence does not contradict the principle of holography of information and, from a fine-grained perspective,  observables in a bounded region are also accessible in its complement.  

Let $|\psi \rangle$ be a state in the bulk theory, with energy that is parametrically larger than the AdS scale 
\be
E = \langle \psi | H | \psi \rangle \gg 1,
\ee
where the AdS scale is set to unity.
In addition, we require the {\em spread} of energies in the state to be large
\be
\delta E = \sqrt{\langle \psi | H^2 | \psi \rangle - \langle \psi | H | \psi \rangle^2} \gg 1,
\ee
while ${\delta E \over E} \ll 1$. 

In a typical state with these properties, it is possible to ``dress'' operators so that they approximately commute with the ADM Hamiltonian.  Let $\alset$ be the set of simple bulk operators that we are interested in. For instance, this set might correspond to products of bulk field operators in some region
\be
\label{alsetdef}
\alset = \text{span}\{\phi(x_1), \phi(x_1) \phi(x_2), \ldots \phi(x_1) \phi(x_2) \ldots \phi(x_n) \}.
\ee
Two points must be kept in mind.
\begin{enumerate}
\item
The Hamiltonian itself is not included in $\alset$. The distinction between ordinary operators and the Hamiltonian is clear at lowest order in perturbation theory but subtle beyond lowest-order since the OPE of ordinary operators can generate the Hamiltonian. 
\item
This set is not quite an algebra since the order of the highest monomial that appears in \eqref{alsetdef} must be cut off to prevent the construction from breaking down. We refer the reader to \cite{Papadodimas:2013jku,Papadodimas:2015xma,Papadodimas:2015jra,Jensen:2024dnl} for more discussion of the cutoff.
\end{enumerate}

We may now consider a family of ```little Hilbert spaces'' \cite{Papadodimas:2013jku} produced by first time-translating  $|\psi \rangle$  and then acting with elements of $\alset$ on $|\psi \rangle$
\be
{\cal H}_{t} = \alset e^{-i H t} |\psi \rangle.
\ee
Under reasonable conditions (see \cite{Jensen:2024dnl} for a more detailed discussion) it may be shown that ${\cal H}_{t}$ is dense in the space
\be
{\cal H}_{t}' =  e^{-i H t} \alset |\psi \rangle.
\ee
This is a limited version of the Reeh-Schlieder property and we will use this property below.

For a typical heavy state we expect that correlators of simple operators are time translationally invariant.
\be
\langle \psi | e^{i H t} a e^{-i H t} | \psi \rangle = \langle \psi | a | \psi \rangle.
\ee
For a typical state, $|\psi \rangle$ one may also show that these little Hilbert spaces are orthogonal to each other if the difference in time-parameters is larger than ${1 \over \delta E}$.\footnote{As discussed in \cite{Jensen:2024dnl}, this assumes that functions of $H$ are not elements of $\alset$. For instance $\langle \psi | {e^{-i H t} \over H} | \psi \rangle = \langle \psi | 1 - i \int_0^{t} e^{-i H s} d s | \psi \rangle \neq 0$. Therefore, at higher-orders in perturbation theory, where $H$ can be generated through OPEs of other operators, the construction needs to be refined.} 
\be
|\langle \psi | a e^{-i H t} | \psi \rangle | \ll 1, \qquad \forall ~a ~\in~\alset,~\text{if}~|t| \gg {1 \over \delta E}.
\ee

Now, let $\delta T$ be a time interval with the property that
\be
1 \gg {\delta T} \gg {1 \over \delta E}.
\ee
We now define a {\em smeared} Hilbert space 
\be
{\cal H}^{\text{sm}}_{t} = \text{span}\{|v \rangle: \quad |v \rangle \in {\cal H}_{s}~\text{for}~s \in (t - {\delta T \over 2}, t + {\delta T \over 2}) \},
\ee
and we define $P_{t}$ to be the projector onto ${\cal H}^{\text{sm}}_t$. Alternately we can think of $P_t$ as the smallest projector that contains projectors onto all the spaces ${\cal H}_s$ for $s \in (t - {\delta T \over 2}, t + {\delta T \over 2})$.

The smearing above is physically motivated and was first introduced in \cite{Jensen:2024dnl}. It causes the relational operators to be defined below to be spread out over a time interval of size ${\delta T}$. Such a spread is to be expected since an observer with energy spread ${\delta E}$ can only measure time intervals larger than ${1 \over \delta E}$.  This smearing was not utilized in the original construction of \cite{Papadodimas:2015xma,Papadodimas:2015jra}, where projectors onto a specific ${\cal H}_t$ were used instead; this oversight was carried over to  \cite{Bahiru:2023zlc,Bahiru:2022oas}. However, projectors onto a specific ${\cal H}_t$ are ``too sharp'' and, we will see below, that they cause difficulty when one studies the action of relational-operators on the Hilbert space or higher-point function of relational operators. 

The original elements of $\alset$, in general, do not commute with $H$. However, one can dress them to obtain operators that do commute with $H$ to high precision.
\be
\label{hatadef}
\hat{a} = {1 \over \delta T} \int_{-T_{\text{cut}}}^{T_{\text{cut}}} e^{i H t} a P_0 e^{-i H t} dt.
\ee
We refer the reader to \cite{Papadodimas:2015jra} for extensive discussion of the cutoff $T_{\text{cut}}$ and its relation to state dependence. 

However, within low-point correlators about the state $|\psi \rangle$ and nearby time-translated states the precise value of $T_{\text{cut}}$ is unimportant since the state $e^{-i H t} |\psi \rangle$ is orthogonal to ${\cal H}^{\text{sm}}_{0}$. For any $b,c \in \alset$,
\be
\langle \psi | c {\partial \hat{a} \over \partial T_{\text{cut}}} b | \psi \rangle \approx 0.
\ee
Hence, for the purpose of the current discussion, we can simply assume that $T_{\text{cut}}$ is effectively infinite.

Up to the approximations enumerated explicitly above, these operators commute with the Hamiltonian to excellent accuracy
\be
e^{i H s} \hat{a} e^{-i H s} = \hat{a},
\ee
since the only effect of the time translation is to change the limit of the integral over $t$ in \eqref{hatadef} which, as explained above,  is irrelevant in low-point correlators.

The action of these operators on any element of the original little Hilbert space is given by
\be
\label{hataonbpsi}
\hat{a} b |\psi \rangle = {1 \over \delta T} \int_{-{\delta T \over 2}}^{{\delta T \over 2}} e^{i H t} a e^{-i H t} d t b |\psi \rangle + \Or[{1 \over \delta E \delta T}] \approx \bar{a} b |\psi \rangle,
\ee
where $\bar{a}$ is a smeared version of the original operator obtained by time averaging it over an interval of size $\delta T$.
\be
\bar{a} \equiv {1 \over \delta T} \int_{-{\delta T \over 2}}^{{\delta T \over 2}} e^{i H t} a e^{-i H t} d t.
\ee
The error term above arises because we recall that projectors onto different little Hilbert spaces are not exactly orthogonal. So the $P_0$ projector also picks up contributions from the $t$ integral in the range $|t| \in ({\delta T \over 2}, {\delta T \over 2} + {1 \over \delta E})$. However, in the regime where $\delta T \gg {1 \over \delta E}$, this error is negligible. Conversely, if one chooses $\delta T$ to be too small, it leads to large errors above and the action of the hatted operators is no longer close to the action of the original operators.\footnote{As noted, the construction \eqref{hatadef} is a slight variant of mirror-operator construction. Mirror operators are usually constructed \cite{Papadodimas:2013jku,Papadodimas:2013wnh} by specifying their action on a state through relations of the form \eqref{hataonbpsi}. The projector construction was presented in \cite{Papadodimas:2015jra}  as an alternative to a parallel construction that defined the mirror operators in the eternal black hole in terms of their action on states; this parallel construction is also described in \cite{Bahiru:2023zlc}. But unless one smears the projectors appropriately, the two constructions are not equivalent. Even without smearing, the one-point function of the operators works out correctly as shown in section 7.6 of \cite{Papadodimas:2015jra}; however,  the action on the Hilbert space is not accurate and this affects higher-point functions as was also recently noted in \cite{Antonini:2025sur}. However \cite{Antonini:2025sur} propose a fine-tuned modification of the projector whereas we feel that the smearing, as earlier described in \cite{Jensen:2024dnl}, provides a more physical solution to the issue.}

Note that $\bar{a} b |\psi \rangle$ is also an element of ${\cal H}^0$ by the Reeh-Schlieder property alluded to above. A short calculation then shows that the correlators of the hatted operators in the original state reduce to correlators of averaged operators
\be
\langle \psi | \hat{a}_1 \ldots \hat{a}_n | \psi \rangle = \langle \psi | \bar{a}_1 \ldots \bar{a}_n | \psi \rangle.
\ee

To the extent that $\delta T$ is small compared to other time intervals in the problem the distinction between $\bar{a}$ and $a$  is physically irrelevant. Therefore, in this regime the hatted operators approximately commute with the Hamiltonian and act approximately like the original operators within low-point correlators. They may be thought of as relational operators that are measured by an ``observer'' with an autonomous clock derived from the properties of the state $|\psi \rangle$ that is distinct from the clock on the boundary.
\subsection{Observer states \label{subsecobs}}
In this appendix, we provide an explicit example of a free-field state that satisfies the conditions required of a background QFT state in Section \ref{secrelational}.

Let $\phi(x)$ be the late-time limit of a principal series field. We use this notation to indicate that the background state might comprise a field that is distinct from the observable field, $\chi(x)$ although the construction below allows $\phi = \chi$.

Now consider, 
\be
|\psiback \rangle =  \int \left( \prod_{i=1}^{S}d x_i f_i(x_i) \right) : \phi(x_1)  \ldots \phi(x_S): | 0 \rangle,
\ee
where the functions $f_1(x_1) \ldots f_S(x_S)$ are some $L^2$ normalizable smearing functions that have compact support in the region $R'$ and the product of fields is normal ordered. We will consider the limit where $S \gg 1$, so that the state contains a ``large number'' of particles.

We want these smearing functions to be ``random''. The space of $L^2$ normalizable functions is infinite-dimensional but we can regularize it to avoid functions that vary too rapidly and then use the Haar measure on this space to select the smearing functions $f_i$.  We then expect these smearing functions to be uncorrelated under the standard $L^2$ inner product. For $i \neq j$,
\be
\langle f_i, f_j \rangle = \int f_i^*(x) f_j(x) d x = 0,
\ee
to a good approximation. More precisely if we regularize the space of $L^2$-normalizable functions to have dimension $D$, and each smearing function $f_i$ to have unit norm,  the typical size of the inner product is expected to be ${1 \over \sqrt{D}}$

The normalization constant, ${\cal N}_2$ can then be chosen as 
\be
{\cal N}_2^{-2} = \langle \psiback |\psiback \rangle =   \prod_i \mathfrak{c}_{\phi \bar{\phi}} \langle f_i, f_i \rangle,
\ee
so the state is unit normalized. Here, the normalization of the propagator is the same as \eqref{allprops}

Now consider an element of the dS group that maps point $x_i \rightarrow \tilde{x}_i$. This acts on the state as
\be
U |\psiback \rangle = {\cal N}_2 \int \left(\prod_{i=1}^{S} d x_i \tilde{f}(x_i) \right) : \phi(x_1) \ldots \phi(x_S): | 0 \rangle,
\ee
where 
\be
\tilde{f}(x_i) = |{\partial \tilde{x}_i \over \partial x_i}|^{\bar{\Delta} \over d} f(\tilde{x}_i).
\ee

When $U$ is close to the identity so that the points are moved by only a small amount, we see that
\be
\label{innerprod}
\langle \psiback | U |\psiback \rangle = \prod_i { \langle f_i, \tilde{f}_i \rangle \over \sqrt{\langle f_i, f_i \rangle \langle \tilde{f}_i, \tilde{f}_i} \rangle } \equiv \prod_i (1 - \delta_i),
\ee
where the numbers $\delta_i$ are defined by the relation above. Here, we have neglected overlaps $\langle f_i, \tilde{f}_j \rangle$ that come from other Wick contractions since we do not expect a generic conformal transformation to generate correlations between these random smearing functions.

In the case where we have a large number of field insertions, we can approximate
\be
\prod_i (1 - \delta_i) \approx e^{-S \delta} \ll 1,
\ee
where $\delta$ is the ``mean value'' of the $\delta_i$. So if the transformation $U$ moves points by an amount that is larger than the typical scale of variation of the $f_i$, we expect that $\delta = \Or[1]$ and the inner product then drops off rapidly if $S$ is large. 

We see that this state satisfies both the necessary conditions outlined in the main text. Since the insertions of $\phi$ are localized in a region that is well separated from the region $R$, we see that
\be
\label{wellsep}
\langle \psiback | a |\psiback \rangle = \langle a \rangle.
\ee
Second, 
\be
\label{sharplylocal}
\langle \psiback | U a | \psiback \rangle \ll 1
\ee
for $U$ outside of a small neighbourhood of the identity.

While an explicit construction is difficult to present in the interacting theory, we expect that a state that satisfies the properties above will continue to exist in the interacting theory.  First,  \eqref{wellsep} is expected to hold on the basis of cluster decomposition if the excitation and the observables are located on very different parts of the late-time slice.  Second, we used the free-field two-point function to show \eqref{sharplylocal}. In the interacting theory, the norm does not factorize into a product of two-point functions.  While perturbative corrections can remove some of the factors of $(1 - \delta_i)$ in \eqref{innerprod}, this does not alter \eqref{sharplylocal} unless we go to S\textsuperscript{th} order in perturbation theory. But such a contribution is highly suppressed in any case.

\section{Wavefunctions for the principal series \label{appprincipalwavefunc}}
In this Appendix, we quickly review how principal-series matter fields in de Sitter can be treated in a wavefunction formalism. This issue was clarified in \cite{Dey:2024zjx}. Principal-series wavefunctions in de Sitter space are essentially coherent-state wavefunctions. Before we proceed to the de Sitter case, we remind the reader of such wavefunctions in ordinary quantum mechanics; the discussion for de Sitter field theory will be parallel to this.

\subsection{Coherent state wavefunctions in quantum mechanics}
Consider a single particle with a Hamiltonian 
\be
H_{1p} = {p^2 \over 2} + {x^2 \over 2 }.
\ee
The vacuum is annihilated by the standard annihilation operator
\be
A = (x + i p)/\sqrt{2}.
\ee

To bring out the analogy with de Sitter space, we define new creation and annihilation operators
\be
a = \alpha A + \beta A^{\dagger}; a^{\dagger} = \beta^* A + \alpha^* A^{\dagger},
\ee
where $\alpha, \beta$ are complex numbers that satisfy $|\alpha|^2 - |\beta|^2 = 1$. These operators do not annihilate the usual vacuum. Instead we can define a different state $|\tilde{0} \rangle$ that satisfies $a |\tilde{0} \rangle = 0$ and is normalized so that $\langle \tilde{0} | \tilde{0} \rangle = 1$.

We now define the state $| z\rangle$ to satisfy
\be
a | z \rangle = | z \rangle,
\ee
which fixes the state up to normalization to be
\be
\label{zasexcit}
|z \rangle =  e^{z a^{\dagger}} | \tilde{0} \rangle.
\ee

It is not difficult to check that these states obey the completeness relation
\be
\int [d^2 z] \, | z \rangle \langle z | = 1,
\ee
where $[d^2 z] \equiv e^{-|z|^2} {d^2 z \over \pi} $. 
This identity is proved by expanding the states \eqref{zasexcit} in terms of normalized eigenstates of the number operator $a^{\dagger} a |n \rangle = n | n \rangle$. 
\be
|z \rangle =  \sum_n {z^n \over \sqrt{n!}} |n \rangle,
\ee
and then noting that
\be
\int {d^2 z  \over \pi} \sum_{n,m} e^{-|z|^2} z^n \bar{z}^m |n \rangle \langle m| = \sum_{n} |n \rangle \langle n | = 1.
\ee

Now, for any state, $| \Psi \rangle$, we can define the wavefunction
\be
\Psi(z) = \langle z | \Psi \rangle.
\ee
This notation is slightly deceptive since $\Psi(z)$ is, in general, not a holomorphic function of $z$. 
We can compute expectation values of polynomials of $z$ and $\bar{z}$ by integrating the polynomial against $|\Psi(z)|^2$. In the operator formalism, such an integral yields expectation values where the $a^{\dagger}$ appears to the right of the $a$.
\be
\int [d^2 z] \,  |\Psi(z)|^2 z^n \bar{z}^m = \int [d^2 z] \, \langle \Psi | a^n | z \rangle \langle z | (a^{\dagger})^m | \Psi \rangle = \langle \Psi | a^n (a^{\dagger})^m | \Psi \rangle.
\ee
Here, we first used the fact that $|z \rangle$ is a right eigenvector of $a$ and $\langle z|$ is a left eigenvector of $a^{\dagger}$ followed by the use of the completeness relation.
 
The analysis so far was kinematic and independent of the Hamiltonian.  Now, we turn to the wavefunction of the true ground state
\be
\Psi_0(z) = \langle z | 0 \rangle.
\ee
Since $A = \alpha^* a - \beta a^{\dagger}$, and $A |0 \rangle = 0$, one can write
\be
|0 \rangle = N e^{ {\beta \over 2 \alpha^*} (a^{\dagger})^2} |\tilde{0} \rangle,
\ee
where $N$ is a normalization we will fix below. Since $\langle z|$ is a left eigenvector of $a^{\dagger}$ we find
\be
\langle z | 0 \rangle = N e^{ {\beta \over 2 \alpha^*} \bar{z}^2}.
\ee
The normalization can be fixed by demanding 
\be
\int |\Psi_0(z)|^2 {[d^2 z]} = 1,
\ee
which fixes
\be
N = {1 \over |\alpha|^{1 \over 2}}.
\ee

The wavefunction for states obtained by acting with powers $a^{\dagger}$ on the true vacuum is
\be
\langle z| (a^{\dagger})^m | 0 \rangle = \bar{z}^m \Psi_{0}(z)= {1 \over |\alpha|^{1 \over 2}} \bar{z}^m e^{\bar{z}^2 {\beta \over 2 \alpha^*}}.
\ee
We see that these wavefunctions are all  antiholomorphic. This set of antiholomorphic wavefunctions already yield a complete basis of states.

There is nothing wrong with wavefunctions that depend on $z$ rather than $\bar{z}$. For instance the norm of the wavefunction $z \Psi_{0}(z)$ is the same as the norm of $\bar{z} \Psi_0(z)$. However, the states corresponding
to these wavefunctions do not have a simple operator representation in terms of the action of $a$ or $a^{\dagger}$ on $|0 \rangle$. Since we already have a complete basis above, it is redundant to study such wavefunctions.

\subsection{Wavefunctions for principal-series fields}
In \cite{Chakraborty:2023yed}, the states were described as joint-wavefunctions of metric and matter fluctuations. As explained in the main text, strictly speaking,  one must mod-square this joint wavefunction and then integrate out the metric fluctuations in order to focus on the matter fluctuations. In this Appendix, we provide a simplified treatment starting with quantum-field theory in de Sitter space where we simply neglect the metric fluctuations from the start. We then group average these wavefunctions to obtain effective wavefunctions that can be used to compute gravitational correlators in the $G_N \to 0$ limit.

The canonical commutation relation of the bulk field with its conjugate momentum imply that the boundary values, $\chi(x)$ and $\bchi(x)$ satisfy
\be\label{eq:chi_comm_rel}
	[\chi(x),\bchi(x')] = \frac{1}{2\mu}\delta(x-x'),
\ee
where $\Delta = {d \over 2} + i \mu$. 

Precisely as in the quantum-mechanical example above, one can define coherent states that satisfy
\be
\chi(x) |\chi \rangle = \chi(x) |\chi \rangle.
\ee
Here we have abused notation since the left hand side displays the operator $\chi(x)$ acting on a state and the right hand side displays an ordinary c-number function $\chi(x)$ acting on the same state. We hope that the context will always make it clear when $\chi(x)$ is used as an operator and when it is used as an ordinary function. We normalize these states, just as in the quantum-mechanical example, using
\be
\label{normchiwave}
\langle \chi  | \chi \rangle = e^{2 \mu \int d x |\chi(x)|^2},
\ee
so that they satisfy the completeness relation
\be
\label{chinormchoice}
\int [D \chi D \bchi] |\chi \rangle \langle \chi | = 1,
\ee
where 
\be
[D \chi D \bchi] \equiv D \chi D \bchi e^{-2 \mu \int d x |\chi(x)|^2},
\ee
and $D \chi D \bchi$ is the usual functional measure.

As is well known the Euclidean vacuum itself is not annihilated by $\chi(x)$ \cite{Mottola:1984ar}. Instead, the correct positive frequency modes are those that are well behaved when one continues global de Sitter time, $\tau$ to the complex plane and studies the limit $\tau \rightarrow -i {\pi \over 2}$; the negative-frequency modes are those that are well behaved in the limit $\tau \rightarrow +i {\pi \over 2}$. This is precisely the analogue of the quantum-mechanical problem studied above. The wavefunction of the free Euclidean vacuum is given by \cite{Dey:2024zjx}
\be\label{eq:euc_wavefunc}
	\Psi_0[\chi] :=\braket{\chi|0} = \exp\encbr{\frac{\mu\b^*}{\a^* C} \int  dx_1  dx_2  \frac{\bchi(x_1)\bchi(x_2)}{\abs{x_1-x_2}^{2\D}}}.
\ee
\be
	\frac{\b}{\a} = \frac{2^{2i\mu }\mu \Gamma (i\mu )^2}{\pi\enc{1+\coth(\pi\mu)}}, \qquad
	C =  \frac{2^{d-2\Delta} \pi^{d/2} \Gamma (d/2-\Delta)}{\Gamma(\Delta)}.
\ee
This is consistent with the momentum space expression in (5.21) of \cite{Dey:2024zjx}, after we recall the relation 
\be
{\abs{x}}^{-2\D} = C\int \frac{dp}{(2\pi)^d} e^{-ix\cdot p} \abs{p}^{2\D-d},
\ee
with $C$ as above. The expression in \cite{Dey:2024zjx} is valid for all times and also includes an additional term in the exponent that is proportional to $(-\eta)^{i \mu}$ where $\eta$ is the Poincare time. Here, it is removed by the choice of normalization \eqref{normchiwave} and \eqref{chinormchoice} and by discarding a phase that is not relevant for our computations.

In the presence of interactions, the Euclidean vacuum is no longer expected to be Gaussian in the fields and instead we have
\be\label{eq:bd_heavy_form}
	\Psi_0 [\chi] = \exp \encbr{\sum_{n=2}^\infty \int  dx_1\dots  dx_n\,G_n(x_1,\dots,x_n) \bchi(x_1)\dots\bchi(x_n)},
\ee
where $G_n$ satisfy the Ward identities of connected conformal correlators of operators with dimension $\dbar$. 
Excited dS-invariant states take on the form
\be
	\Psi[\chi] = \int  dx_1\dots  dx_m\, \dcoeff[m] \bchi(x_1)\dots\bchi(x_m) \Psi_0[\chi],
\ee
where $\dcoeff[m]$ is the difference of two possible $G_m$. 
In ket notation, this is
\be
	\ket{\Psi} = \int  dx_1\dots  dx_m\, \dcoeff[m](x_1,\dots,x_n) \bchi(x_1)\dots\bchi(x_m) \ket{0}.
\ee
As in the quantum-mechanical example studied above, the correspondence between the operator notation and wavefunctions is simplest when we study excitations produced by $\bchi$ that are represented by anti-holomorphic wavefunctions.

Moreover, using the same arguments as in the quantum-mechanical case, integrating the squared wavefunction with a polynomial of $\bchi$ and $\chi$ produces an expectation value where insertions of  $\bchi$ are ordered to the right of insertions of $\chi$.
\be
	\langle \Psi | \chi(x_1)\dots\chi(x_m)\bchi(y_1)\dots \bchi(y_n) |\Psi \rangle  = \int [D\chi D\bchi] \,\abs{\Psi[\chi]}^2 \chi(x_1)\dots\chi(x_m)\bchi(y_1)\dots \bchi(y_n).
\ee

If $\Psi[\chi]$ has non-Gaussian terms in the exponent apart from excitations, these can simply be perturbatively-expanded within the functional integral in the usual manner reducing the entire expression to a sum of
Gaussian expectation values.

The contact term in \eqref{allprops} can be understood as coming from the measure $[D \chi D \bchi]$. This naturally yields the propagator with $\bchi$ on the right, which is all we need in this paper. If needed, the propagator with the other ordering can be obtained by adding the commutator \eqref{eq:chi_comm_rel}.
\bibliographystyle{utphys}
\bibliography{references}

\end{document}